\documentclass[aip,rsi,reprint,graphicx]{revtex4-1}

\usepackage{latexsym}
\usepackage{graphicx}
\usepackage{dcolumn}
\usepackage{bm}
\usepackage{amssymb}

\begin{document}

\preprint{Draft RSI Version 14 (all)}

\title{The Polarized H and D Atomic Beam Source for
ANKE at COSY-J\"ulich}

\thanks{Work financially supported by German Ministry
for Education and Research (BMBF) under contract Nos. RUS-649-96 and 06 ER 831, by
Forschungszentrum J\"ulich (FF\&E) under contract No. 41149451, by Deutsche Forschungsgemeinschaft
under contract No. 436 RUS 113/430, and by the Russian Ministry of Sciences}

\author{M.\,Mikirtychyants}
\email{m.mikirtychyants@fz-juelich.de}
\thanks{Now at Institut f\"ur Experimentalphysik, Ruhr-Universit\"at Bochum,
44801 Bochum, Germany}
\affiliation{Institut f\"ur Kernphysik, Forschungszentrum J\"ulich,
52425 J\"ulich, Germany}
\affiliation{High Energy Physics Department, St.Petersburg Nuclear Physics Institute,
188300 Gatchina, Russia}
\author{R.\,Engels}
\affiliation{Institut f\"ur Kernphysik, Forschungszentrum J\"ulich,
52425 J\"ulich, Germany}
\author{K.\,Grigoryev}
\affiliation{Institut f\"ur Kernphysik, Forschungszentrum J\"ulich,
52425 J\"ulich, Germany}
\affiliation{High Energy Physics Department, St.Petersburg Nuclear Physics Institute,
188300 Gatchina, Russia}
\author{H.\,Kleines}
\affiliation{Zentrallabor f\"ur Elektronik, Forschungszentrum J\"ulich,
52425 J\"ulich, Germany}
\author{P.\,Kravtsov}
\affiliation{High Energy Physics Department, St.Petersburg Nuclear Physics Institute,
188300 Gatchina, Russia}
\author{S.\,Lorenz}
\thanks{Now at Osram GmbH, 93049 Regensburg, Germany}
\affiliation{Physikalisches Institut, Friedrich-Alexander-Universit\"at Erlangen-N\"urnberg,
91058 Erlangen, Germany}
\author{M.\,Nekipelov}
\thanks{Now at Wissenschaftlich-Technische Ingenieurberatung GmbH, 52428 J\"ulich, Germany}
\affiliation{Institut f\"ur Kernphysik, Forschungszentrum J\"ulich,
52425 J\"ulich, Germany}
\affiliation{High Energy Physics Department, St.Petersburg Nuclear Physics Institute,
188300 Gatchina, Russia}
\author{V.\,Nelyubin}
\thanks{Now at  Department of Physics, University of Virginia, Charlottesville, VA 22904, USA}
\affiliation{High Energy Physics Department, St.Petersburg Nuclear Physics Institute,
188300 Gatchina, Russia}
\author{F.\,Rathmann}
\affiliation{Institut f\"ur Kernphysik, Forschungszentrum J\"ulich,
52425 J\"ulich, Germany}
\author{J.\,Sarkadi}
\affiliation{Institut f\"ur Kernphysik, Forschungszentrum J\"ulich,
52425 J\"ulich, Germany}
\author{H.\,Paetz\,gen.\,Schieck}
\affiliation{Institut f\"ur Kernphysik, Universit\"at zu K\"oln,
50937 K\"oln, Germany}
\author{H.\,Seyfarth}
\affiliation{Institut f\"ur Kernphysik, Forschungszentrum J\"ulich,
52425 J\"ulich, Germany}
\author{E.\,Steffens}
\affiliation{Physikalisches Institut, Friedrich-Alexander-Universit\"at Erlangen-N\"urnberg,
91058 Erlangen, Germany}
\author{H.\,Str\"oher}
\affiliation{Institut f\"ur Kernphysik, Forschungszentrum J\"ulich,
52425 J\"ulich, Germany}
\author{A.\,Vasilyev}
\affiliation{High Energy Physics Department, St.Petersburg Nuclear Physics Institute,
188300 Gatchina, Russia}

\date{\today}
\begin{abstract}
A polarized atomic beam source was developed for the polarized internal storage-cell
gas target at the magnet spectrometer ANKE of COSY-J\"ulich. The intensities of the beams
injected into the storage cell, measured with a compression tube, are $7.5\cdot 10^{16}$
hydrogen atoms/s (two hyperfine states) and $3.9\cdot 10^{16}$ deuterium atoms/s (three
hyperfine states). For the hydrogen beam the achieved vector polarizations are
$p_{\rm z}\approx\pm0.92$. For the deuterium beam, the obtained combinations of vector and
tensor ($p_{\rm zz}$) polarizations are $p_{\rm z}\approx\pm 0.90$ (with a constant
$p_{\rm zz}\approx +0.86$), and $p_{\rm zz}=+0.90$ or $p_{\rm zz}=-1.71$ (both with vanishing
$p_{\rm z}$). The paper includes a detailed technical description of the apparatus and of the
investigations performed during the development.
\end{abstract}
\maketitle

\section{Introduction}
Single-polarized experiments, making use of the polarized proton and deuteron beams of the cooler
storage ring COSY-J\"ulich and unpolarized targets, are extended to double-polarized
studies~\cite{COSY152} by the installation of an internal polarized hydrogen or deuterium
storage-cell gas target. Utilizing pure gas of polarized hydrogen or deuterium, these targets
circumvent the problem of dilution by unpolarized nucleons and they permit fast change of the
polarization direction. In order to compensate for the relatively low areal density, these targets
are placed inside storage rings, where they are traversed by the orbiting beam typically
a million times per second. As conceived already some forty years ago~\cite{Haeberli_1965}, a
substantial enhancement of the areal target-gas density (or luminosity) compared to gas-jet
targets by about two orders of magnitude is achieved, when the polarized atoms are injected into
an open-ended, T-shaped storage cell. A review describing the capabilities of polarized gas-jet
and storage-cell gas targets is found in Ref.~\cite{Steffens+Haeberli_2003}.

The polarized internal target (PIT), developed for the magnet spectrometer
ANKE~\cite{Barsov_et_al_2001} in COSY J\"ulich~\cite{Maier_1997}, consists of the polarized atomic
beam source (ABS), the storage cell~\cite{KG_2012}, and the Lamb-shift polarimeter
(LSP)~\cite{Engels_et_al_2003, Engels_et_al_2005_1}. In the development of the PIT for ANKE the
experience of other groups in the operation of polarized gas targets could be used. Essentially
these are the Madison source~\cite{Wise_et_al_1993}, used by the PINTEX  collaboration at
IUCF~\cite{Dezarn_et_al_1995, Rinckel et al 2000}, and the FILTEX source~\cite{Stock_et_al_1994}
used by the HERMES collaboration at DESY/Hamburg~\cite{Nass_et_al_2003}.

Section~\ref{Sec:II} presents the general layout of the ABS and it describes the major elements,
the pumping system, the layout of the baffles, the dissociator, the nozzle-skimmer setup, the
system of sextupole magnets, the high-frequency transition units, and the slow control system.
Section~\ref{Sec:III} contains studies with  the free hydrogen jet from the nozzle. In Secs. IV to
VII measurements and results are  presented concerning the properties of the beam behind the last
magnet, the hydrogen and deuterium beam intensities (\ref{Sec:IV}), hydrogen beam profiles
(\ref{Sec:V}), the degree of dissociation of the hydrogen beam (\ref{Sec:VI}), and the
polarization values for the hydrogen and deuterium beam (\ref{Sec:VII}). Finally, in
Sec.~\ref{Sec:VIII}, the conclusions and an outlook to the future efforts are given. The procedure
of discharge-tube and nozzle conditioning is described in Appendix~\ref{Sec:Appendix}.

\section{Description of the Setup\label{Sec:II}}
\subsection{General layout of the ABS}
The limited space at the target position between the first beam-bending dipole magnet D1 and the
central spectro\-meter dipole magnet D2 of the ANKE facility~\cite{Barsov_et_al_2001} on the one
hand enforces the ABS to be mounted in a vertical position contrary to the layout of the earlier
sources. On the other hand, this position allows less complicated supports of the components and
easier alignment. Furthermore, the height of the COSY tunnel restricts the vertical dimension of
the ABS, which requires a very compact layout of the source. Initially, a slight inclination from
the vertical ABS orientation had been foreseen to avoid drizzling of powder, created during
operation of the dissociator, down into the storage cell. Production of such a powder had been
observed earlier ($\rm{SiO_2}$~\cite{Derenchuk_et_al_1994,Okamura_et_al_1994}, ''green
powder''~\cite{Hatanaka_et_al_1997}). No such powder, however, was found on a glass window at the
exit of the ABS in vertical position after weeks of dissociator operation. Thus, the inclination
could be regarded as unnecessary.

\begin{figure}[!h]
\begin{center}
\includegraphics[width=5.6cm]{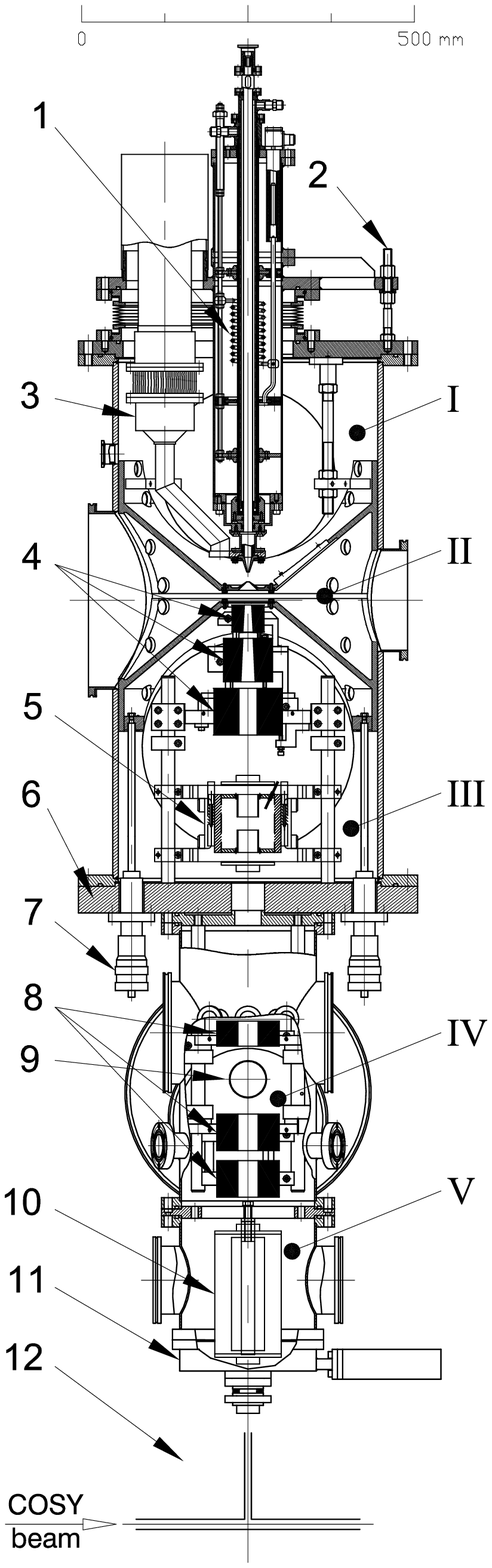}
\caption{Cut along the ABS axis (1: dissociator, 2: one of three adjustment screws for nozzle
positioning, 3: Cu heat-bridge for nozzle cooling with a flexible Cu strands connection to the
coldhead, 4: first set of sextupole magnets, 5: medium field rf transition unit, 6: central
support plate, 7: one of two rotational feed-throughs enabling shift of the lower baffle,
8: second set of sextupole magnets, 9: rotating beam chopper, 10: weak and strong field rf
transition units, 11: vacuum gate valve between ABS and ANKE target chamber, 12: schematical
indication of the storage cell). The labels I to V denote the chambers of the differential
pumping system.}
\label{ABS}
\end{center}
\end{figure}

The layout of the ABS is presented in Fig.~\ref{ABS}. The two main cylindrical vacuum vessels are
fixed above and below a central support plate. The inner diameter of the upper
vessel~\cite{Schiffer} is about
390\,mm. It houses the chambers I, II, and III, which are separated by two conical baffles. The
dissociator tube and the coldhead~\cite{RGS120} with the heat bridge for
nozzle cooling are mounted on a three-legged plate. Three screws allow one to adjust the radial
nozzle position and the distance to the skimmer.

The upper baffle between chamber I and II with the beam skimmer is rigidly connected to the upper
flange of the vessel. The lower baffle with the diaphragm in front of the first sextupole magnet,
separating the chambers II and III, can be moved axially from outside to optimize the pumping
conditions on the one hand between the skimmer and the diaphragm, and on the other in the narrow
region between the diaphragm and the front face of the first magnet. The first set of three
sextupole magnets and the medium field rf transition (MFT) unit in chamber III are carried by two
rods attached to the central support plate. The lower vacuum vessel, chamber IV, has a smaller
inner diameter of 200\,mm. It houses the second set of three sextupole magnets and a cylindrical
beam chopper rotating around a horizontal axis. The separate chamber V provides space for the weak
and strong field rf transition units (WFT and SFT units, respectively). The ABS can be separated
from the ANKE target chamber and the COSY vacuum system by
a gate valve~\cite{VAT-mini} embedded by a dedicated construction into the end flange of chamber
V.

Two strong turbomolecular pumps are installed at opposite flanges of chamber I perpendicular to
the beam direction, one on the beam-up side of chamber II. The chambers III, IV, and V are
evacuated by cryopumps. Due to space limitation around the ABS, shutters on the cryopumps as used
in other sources were omitted. The gas originating from regeneration of the cryopumps is pumped
via bypass tubes by turbomolecular pumps on the upper chambers. During regeneration as in other
cases of pressure increase the gate valve on chamber V is closed to avoid gas flow into the ANKE
target chamber.

Details of the pumping system, the baffles, the dissociator, the area around the nozzle, the
magnet system, the rf transition units,  and the slow control system  are described in the
subsequent subsections.
\subsection{Pumping System}
\begin{figure}[hbt]
\includegraphics[width=7cm]{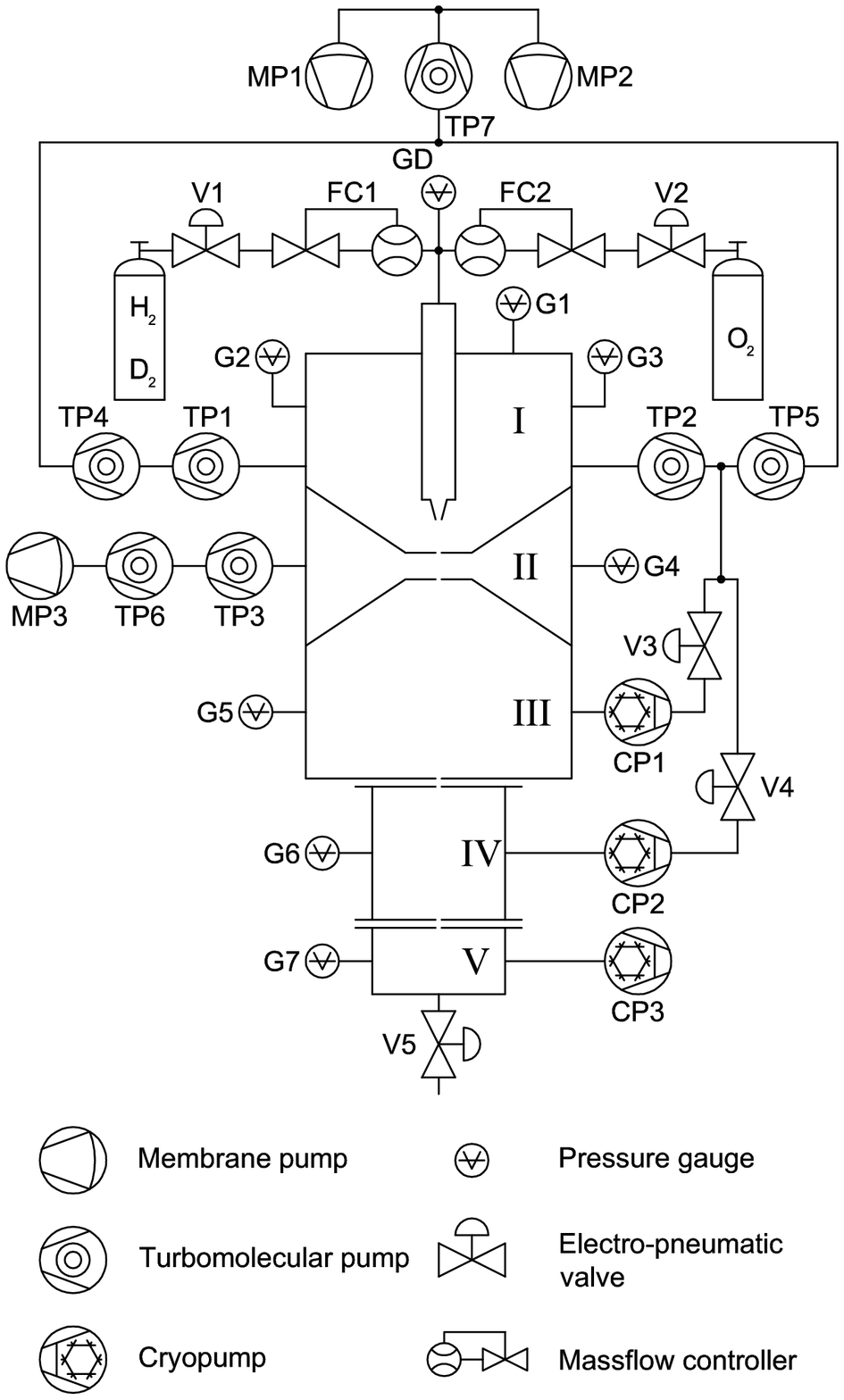}
\caption{The system of pumps on the chambers I to V of the ABS (Fig.~\ref{ABS}). The
specifications of the pumps are listed in Tab.~\ref{Pumps}. The figure also contains
the bypass system for the gas load from regeneration of the cryopumps.}
\label{VacSys}
\end{figure}

The system of pumps on the chambers I to V of the ABS (Fig.~\ref{ABS}) is shown in
Fig.~\ref{VacSys}, the types of the pumps, their pumping speeds,  and the achieved pressures are
listed in Table~\ref{Pumps}. Chamber I with the highest gas load, due to the effect of the
skimmer, is pumped by two strong turbomolecular pumps. Each of them is backed by a smaller
turbomolecular pump.
\begin{table}[b]
\caption{List of the devices employed in the ABS pumping system, composed of turbomolecular pumps
(TP), membrane pumps (MP), and cryopumps (CP), with nominal the individual capacities $C_{\rm
H_{2}}$, the pumping speeds $S_{\rm H_{2}}$, and the achieved pressures at a primary gas flow of
1.0 mbar\,$l$/s.}
\renewcommand*{\arraystretch}{1.2}
\begin{ruledtabular}
\begin{tabular}{lllllr}
Cham- & Device  & Type                           &$C_{\rm H_{2}}$        &$S_{\rm H_{2}}$
& Pressure                \\
ber   &         &                                &[bar\,$l$]     &[$l$/s]       &
[mbar]    \\\hline
I     & TP1-2   &  TPH 2200\tablenotemark[1]     &                       &  2800             &
$10^{-4}$ \\
      & TP4-5,7 &  TMH 260\tablenotemark[1]      &                       &  180 \\\vspace{2mm}
      & MP1-2   &  MVP 100-3\tablenotemark[1]    &                       &
      1.8/1.2\tablenotemark[3]\\
II    & TP3     &  TPH 2200\tablenotemark[1]     &                       &  2800             &
$10^{-6}$  \\
      & TP6     &  TMH 260\tablenotemark[1]      &                       &  180\\\vspace{2mm}
      & MP3     &  MVP 100-3\tablenotemark[1]    &                       &
      1.8/1.2\tablenotemark[3]\\\vspace{2mm}
III   & CP1     &  COOLVAC 3000\tablenotemark[2] &  28\tablenotemark[4]  &  5000             &
$10^{-7}$ \\\vspace{2mm}
IV    & CP2     &  COOLVAC 1500\tablenotemark[2] &  28\tablenotemark[4]  &  5000             &
$5\cdot 10^{-8}$ \\
 V    & CP3     &  COOLVAC 800\tablenotemark[2]  &  4.3\tablenotemark[4] &  1000 &
 $5\cdot10^{-8}$\\
\end{tabular}
\end{ruledtabular}
\tablenotetext[1]{Pfeiffer Vacuum GmbH, 35614 Asslar, Germany.}
\tablenotetext[2]{Leybold Vakuum GmbH, 50968 K\"oln, Germany.}
\tablenotetext[3]{Pumping speed at 1000 mbar/10 mbar}
\tablenotetext[4]{At $10^{-6}$ mbar.}
\label{Pumps}
\end{table}
Their exhausts are connected to a common pump of the same type. The total compression ratios of
the serially connected turbomolecular pumps yields sufficient pumping speed for a
primary molecular gas flow up to 3\,mbar\,$l$/s into the dissociator.
The line of pumps is backed by two oil free membrane pumps. According to the lower gas load,
chamber II is evacuated by a simpler line consisting of two turbomolecular pumps and a membrane
pump. All turbomolecular pumps are operated with use of synthetic oil~\cite{Fomblin}. Compared to mineral oil, synthetic oil allows
longer pumping of hydrogen before oil exchange becomes mandatory. Strong cryo\-pumps are utilized
on chambers III and IV, while the lowest chamber with the WFT and SFT units is evacuated by a
smaller cryopump. All cryopumps are equipped with temperature-controlled heating units for
regeneration on both cooling stages~\cite{HU1}. Heating up to room temperature while pumping the resulting gas load by the
bypass system and  cooling down again takes about 2.5 to 3 hours.
\subsection{Baffles}
\begin{figure}[hbt]
\begin{center}
\includegraphics[width=7cm, angle=-90]{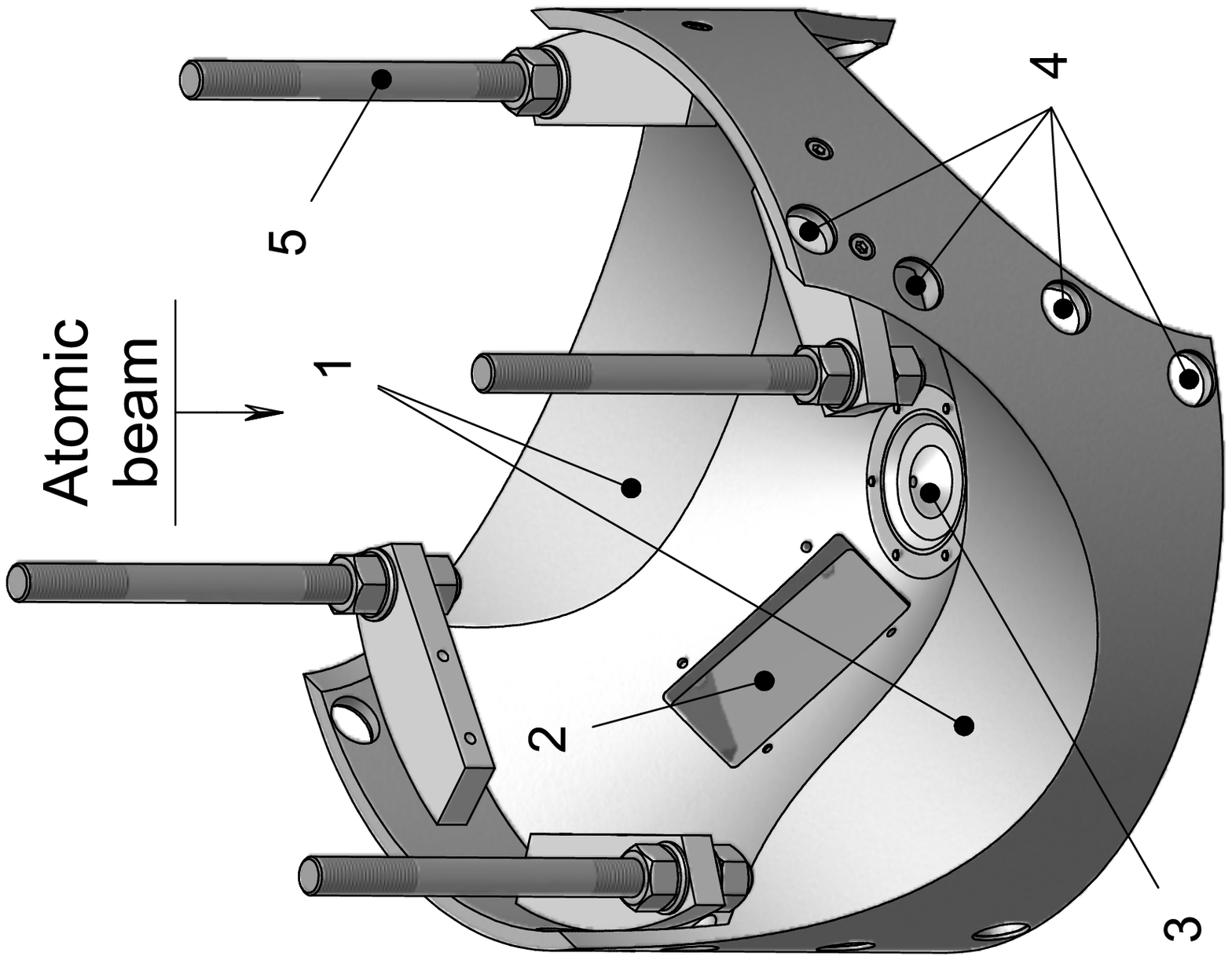}
\caption{3D drawing of the upper baffle, separating the vacuum chambers I and II, with the two
wide cuts in front of the turbopumps (1) and the openings for the viewport (2), the skimmer (3),
four of the 16 ball bearings (4) and the four supporting rods (5).}
\label{baffle}
\end{center}
\end{figure}

The layout of the baffles had been defined by the necessary narrow vertical extension of the ABS
and the requirement to provide sufficient pumping speed in view of the heavy gas load to the
vacuum chambers I and II. Furthermore, the construction aimed at the possibility of axial
movements from outside to optimize the beam parameters by variation of the distances between
nozzle, skimmer, and diaphragm. The resulting shape for the upper baffle is shown in
Fig.~\ref{baffle}. Except for details in the openings, the lower baffle, carrying the diaphragm,
is identical. The layout of the upper vessel and the baffles was done under the boundary
conditions that, on the one hand, the baffles have to be movable within the vessel and, on the
other hand, the slits between cylindrical surfaces of the baffles and the inner surface of the
vessel has to be narrow to reach a small gas conductance. The diameter of the inner vessel
surface is 389.2\,mm with a longitudinal and non-circular tolerance of +0.2\,mm, the outer
diameters of both baffels are (388.7$_{-0.2}$)\,mm. The conductances of the slits of $\le$5\,$l$/s
are small compared with the applied pumping speed. Because of the complicated shape, identical
raw pieces of cast Al~\cite{SKI} were machined to the final dimensions.
Contrary to the lower baffle, the upper baffle until now has to be installed together with the
flange of the upper vessel at a fixed axial position (cf. Fig.~\ref{ABS}). In order to reach full
flexibility in varying the nozzle, skimmer, and collimator relative positions from outside, the
installation of rotational feedthroughs in the flange of the upper vacuum vessel is necessary, a
foreseen but not yet implemented feature.
\subsection{Dissociator}
\begin{figure}[!t]
\begin{center}
\includegraphics[scale=0.7]{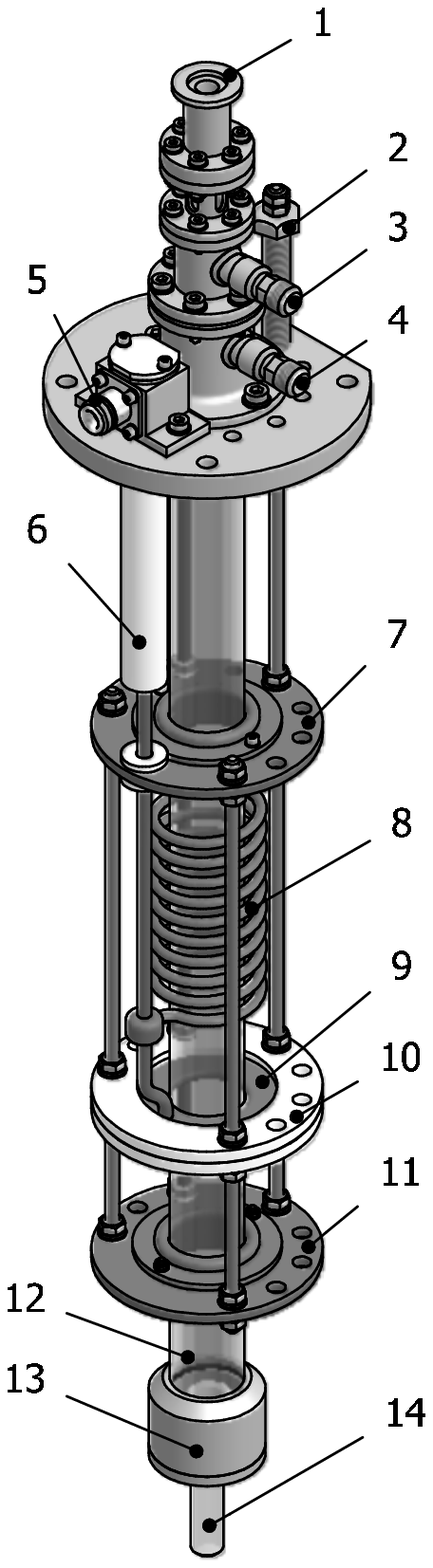}
\caption{3D drawing of the dissociator (1: gas inlet, 2: sliding ground connection, 3: coolant
inlet, 4: coolant outlet, 5: rf input, 6: sliding rf connection, 7: grounded capacitor plate,
8: rf coil, 9: rf-fed capacitor plate, 10: isolating plastic support rings, 11: grounded limiter
plate, 12: lower end of the coolant-guiding tubes, 13: tube support and connection to the coldhead
(details are given in Fig.~\ref{nozzle}), 14: lower end of the discharge tube).}
\label{disso}
\end{center}
\end{figure}
To dissociate molecular hydrogen or deuterium to neutral atoms, an rf discharge is employed which
is fed by a 13.560 MHz generator~\cite{PFG600} delivering up to 600 W into a
50\,$\Omega$ load. The layout of the dissociator, shown in Fig.~\ref{disso}, is similar to that of
the FILTEX design~\cite{Korsch_Diss_1990,Stock_et_al_1996}. The discharge tube ($\varnothing
11\times1.5\,{\rm mm}$)~\cite{TubeDia} is surrounded by two coaxial tubes ($\varnothing 20.4\times1.8\,{\rm mm}$ and
$\varnothing 28\times2\,{\rm mm}$), all three are made from borosilicate glass~\cite{DURAN}. The coolant streams from the inlet
connection down between the discharge tube and the middle tube and, after flow reversal at the
lower end (Fig.~\ref{nozzle}, label 2) it streams up in the outer slit to the outlet connection.
In a closed loop, the coolant inlet temperature (typically 15\,$^{\circ}$C for a 50\% water --
50\% ethanol mixture) is stabilized by a cooling thermostate~\cite{LAUDA}, which would allow
coolant
temperatures down to $-80\,^{\circ}$C. The rf coil and the capacitor, at fixed relative
positions, can be positioned from outside by means of a sliding rf connection~\cite{ODU} and the feed-through ground connection. This enables variation of
the plasma-nozzle distance to optimize the atomic beam intensity while the plasma is burning. The
treatment of the discharge tube and the nozzle prior to installation is described in
Appendix~\ref{sect:appA}.
\subsection{Nozzle}
\begin{figure}[b]
\begin{center}
\includegraphics[scale=0.5, angle=90]{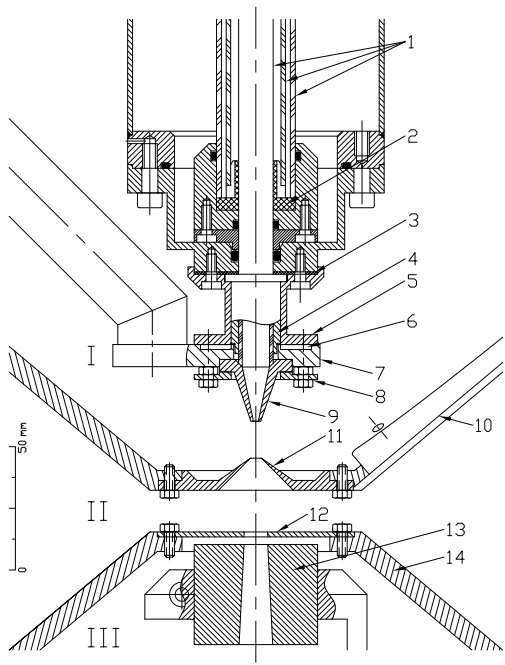}
\caption{Technical drawing including the lower end of the heat bridge and the dissociator, the
nozzle surroundings, and the first sextupole magnet (in scale, 1: discharge and coolant-guiding
tubes, 2: coolant-reversal piece, 3: heat flow reducing Teflon washer, 4: sliding heat connection,
5: stainless steel connector, 6: groove for nozzle-heating element, 7: lower end of the Cu heat
bridge, 8: nozzle fixture, 9: nozzle, 10: baffle, separating the chambers I and II, with a
viewport, 11: stainless steel beam skimmer, 12: Cu diaphragm, 13: first sextupole magnet, and 14:
baffle separating the chambers II and III.}
\label{nozzle}
\end{center}
\end{figure}

The nozzle, cooled via the heat bridge, and the surrounding components are shown in
Fig.~\ref{nozzle}. The nozzle, made from 99.5$\%$ Al, has a simple conical shape with the tip cut.
Comparative measurements show that nozzles with sharp edges as used, e.g., in the Madison
source~\cite{Wise_et_al_1993} do not yield higher atomic beam intensities. First, a sharp edge is
more difficult to produce due to the softness of pure Al. Second, the low heat conductance of a
sharp edge leads to appreciable temperatures of the nozzle tip, caused by recombination of atoms
on the nozzle surface. The temperature at the bottom of the nozzle is measured with a Pt-100
sensor and it is stabilized with an accuracy of $\pm$0.5\,K utilizing a heater. Measurements
with temperature sensors placed along the outer nozzle surface have shown a temperature increase
from 60\,K at the nozzle bottom to $\sim$200\,K at the sharp nozzle tip. In the following, the
nozzle temperature is defined as that measured with this Pt-100 sensor.

With the present system of sextupole magnets, the maximum atomic beam intensity feeding the
storage cell is obtained with a nozzle-orifice diameter of 2.3\,mm and a nozzle-tip to skimmer-tip
distance of 15\,mm at a skimmer-tip diameter of 4.4\,mm and a skimmer-tip to diaphragm distance of
17\,mm. The 2\,mm thick diaphragm with a conical bore, opening from  9.5\,mm to 9.9\,mm towards
the first permanent sextupole magnet, shields the magnet from heating by atoms recombining on its
surface. The slit between the diaphragm and the front face of the magnet enables pumping of gas
from the entrance to the magnet.

The Teflon washer and the stainless steel support separate the cold lower end of the heat bridge
from the warm lower end of the dissociator. The dimensions of these two components  and the\
sliding heat connector, a worked-over sliding high current connector similar to the rf connector
in the dissociator, define the temperature of the lower end of the discharge tube relative to that
of the nozzle. The discharge tube, adapted at its lower end to the nozzle by a chamfered
edge, is pressed to the nozzle by a viton O-ring at its upper end. The two O-rings around
the discharge tube in the lower part of the dissociator seal against the atmosphere. By this
design, only minor forces are exerted to the discharge tube.

The removable viewport in the baffle and the window flange in the upper vacuum vessel (on the
right-hand side of chamber II in Fig.~\ref{ABS}) allows one to observe the nozzle status from the
outside and to exchange nozzles without removal of the dissociator from the setup.

The heat bridge from the coldhead to the nozzle is made from electrolytic Cu. The
flexible link between the coldhead and the heat bridge, consisting of about 200 high-purity Cu
strands of 1 mm diameter, allows for the thermal expansions of the cold and the warm components.
The total cross section of the strands and their heat conductance is smaller than that of a
massive Cu body. This deficiency, however, is reduced by clamping the flexible link directly to
the coldhead. At its operating temperature of about 30\,K, the thermal conductivity of Cu is
about 11, 9, and 5 times higher than that at 300, 100, and 60\,K,
respectively~\cite{Handbuch_Chem_Phys}. Thus, the reduction of the conductance of the entire heat
bridge by the flexible link is minimized by placing it at the coldhead. With the present system,
cooling the nozzle down from room temperature to 60\,K needs about 1.5 hours. The heating element
facilitates warming up to room temperature within about one hour.

Furthermore, avoiding the maze of cold Cu strands around the nozzle, i.e., a labyrinthic cold
surface, compared to an earlier solution~\cite{Koch+Steffens_1999} leads to improved pumping
conditions in the nozzle-skimmer area, where the highest gas load has to be pumped off.

In an earlier phase of the ABS development, attempts have been made to use a cryogenic Ne
heat-pipe of 20\,W cooling power instead of the usual solid Cu bridge to achieve faster cooling and
warming of the nozzle because of the lower heat capacity~\cite{Vassiliev_et_al_1997}. An observed
instability in the necessary operation mode, however, lead to difficulties in nozzle-temperature
stabilization. In view of the fact that the cooling and warming-up times, reached with the Cu
bridge, were satisfying and that its use avoids the additional precautions, imposed by the
heat-pipe operation, it has been replaced by the Cu bridge.
\subsection{Magnet System\label{Sec:II-Magnets}}
The design of the magnet system was made for a set of sextupole magnets consisting of permanently
magnetized segments made from NdFeB compounds, delivering pole-tip fields around 1.5 T. Tracking
calculations from the nozzle to the feeding tube of the storage cell were performed with the use
of a computer code originally developed for the FILTEX ABS~\cite{Korsch_Diss_1990}. The boundary
conditions by the layout of the target setup were the available distance of about 1250\,mm from
the nozzle to the feeding-tube entrance of 10\,mm diameter and the distance from the exit of the
last magnet to the feeding-tube entrance of 300\,mm, necessary to install the SFT and WFT units
and the gate valve between the ABS and the target chamber.

The laboratory velocity distribution of the atoms in the supersonic beam from the nozzle is
described by a modified Maxwellian distribution
\begin{eqnarray}
\mathcal{F}(\vec{v}_{\rm d},T_{\rm b})=\big(\frac{m}{2\,k\,T_{\rm
b}}\big)^{3/2}\exp\left[\frac{-m}{2\,k\,T_{\rm b}}(\vec{v}-\vec{v}_{\rm d})^{2} \right],
\label{velocity-distribution}
\end{eqnarray}
where $m$ is the mass of the atoms and $k$ is the Boltzmann constant. According to time-of-flight
studies~\cite{Lorentz_Dipl_1993}, the drift velocity along the beam axis, $v_{\rm d}$, and the
beam temperature $T_{\rm b}$ for a primary molecular gas flow of 1 mbar\,${\it l}$/s and a
nozzle-orifice diameter of 2\,mm follow a linear dependence on the nozzle temperature $T_{\rm n}$.
For hydrogen $v_{\rm d}[{\rm m/s}]=1351 + 6.1\cdot T_{\rm n}[{\rm K}]$ and $T_{\rm b}=0.29\cdot
T_{\rm n}$ and for deuterium $v_{\rm d}[{\rm m/s}]=1070 + 3.45\cdot T_{\rm n}[{\rm K}]$ and
$T_{\rm b}=0.25\cdot T_{\rm n}$.

As starting conditions of a track a random generator selects a point in the nozzle orifice, one
within the diaphragm in front of the first magnet, and an atom velocity $|v|$. In linear molecular
flow approximation (cf. the discussion in Ref.~\cite{Nass+Steffens_2009}) this defines $\vec{v}$
for the track between the nozzle and the first magnet. According to the geometrical boundary
conditions and the velocity distribution of Eq.~(\ref{velocity-distribution}) the event is either
rejected or used in the further track calculation. Within the magnet the evolution of the track is
calculated stepwise by numerical integration of the equation of motion over integration times of
2\,$\mu s$, corresponding to track lengths of 3.6\,mm for a typical particle velocity of
1800\,m/s. The pure radial force, acting on an atom within the field of the sextupole magnet, is
$\vec{F}_{\rm r}=-\mu_{\rm eff}\cdot\delta B/\delta r\cdot\vec{r}/r$. The effective magnetic
moment, resulting from the Breit-Rabi diagram (e.g., Ref.~\cite{Haeberli_1967}) as
$\mu_{\rm eff}=\delta W /\delta B$, is positive (negative) for atoms in the hyperfine states with
the electron spin parallel (antiparallel) to $\vec{B}$ in the magnet aperture which therefore are
deflected towards (away from) the beam axis. In the drift sections between the two magnet groups
and between the last magnet and the feeding tube the trajectories are assumed as straight lines.

\begin{figure}[b]
\includegraphics[width=\columnwidth]{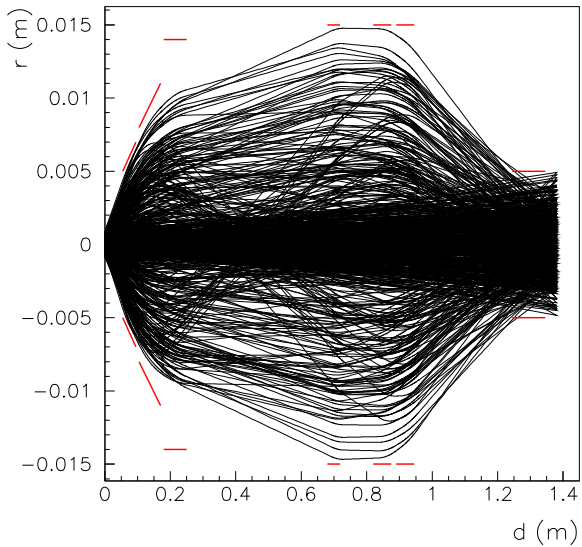}
\caption{Projection of the 3-dimensional trajectories of hydrogen atoms in hyperfine states
$|1\rangle$ and $|2\rangle$ (effective magnetic moment $\mu_{\rm eff}>0)$) from the nozzle
($\varnothing=2\,{\rm mm}, T_{\rm n}=60\,{\rm K})$ to the storage cell calculated for the magnet
arrangement of Table~\ref{magnet-system} and pole-tip fields of 1.5\,T. The positions and lateral
dimensions of the six magnets and the feeding tube are indicated (in red).}
\label{trajectories}
\end{figure}

A variety of systems were studied, all under the assumption of $T_{\rm n}=60\,\rm{K}$ and pole-tip
fields of 1.5\,T. A system utilizing 6 magnets was found to yield satisfying both separation of
the atoms in the $\mu_{\rm eff}<0$ and $\mu_{\rm eff}>0$ states and focusing of the
$\mu_{\rm eff}>0$ states into the feeding tube. Optimization of the parameters led to the system
listed in Table~\ref{magnet-system}. (The tracking calculations, yielding the magnet dimensions
for the order to the manufacturer had been performed for a slightly different geometry.) The table
gives the two distances, at which intensity measurements with the compression tube were performed.
The Fig.~\ref{trajectories} shows the projection of the trajectories of H atoms in the
$\mu_{\rm eff}>0$ states, calculated for this system. One recognizes two groups of trajectories,
one with an intermediate focus and another one with focusing into the feeding tube. The present
result like those of other groups (see e.g., Ref.~\cite{Lorentz_Dipl_1993}) confirms the
expectation~\cite{Kubischta_1991} that the transmission as function of the atom velocity should
show two maxima, one below and one above the most probable velocity.
\begin{table}[t]
\caption{Final dimensions and axial positions of the source components (pole-tip field strenghts
$B^{*}_{0}$ as measured after delivery \cite{Vassiliev_et_al_2000}, inner diameters
($\varnothing_0$), outer diameters ($\varnothing_1$), axial dimensions ($\ell$), and distances
($\Delta$) between the components. The lower three lines give the two distances and the dimensions
of the compression tube used in the intensity measurements.}
\begin{ruledtabular}
\begin{tabular}{cccccc}
component & $B^*_0$ [T] & $\varnothing_0$ [mm] & $\varnothing_1$ [mm] & $\ell$ [mm] & $\Delta$
[mm]\\
\hline &&&&&\\
Nozzle orifice   &        & 2.3      & 3.3   &           &\\
                 &        &          &       &           & 15.0\\
Skimmer          &        & 4.4/30.4\footnote{Conical opening,the first number denotes the
measured diameter of the  entrance, the second that of the exit aperture.}
                                     &       & 13.0      &\\
                 &        &          &       &           & 16.9\\
Diaphragm        &        & 9.5/9.9\footnotemark[1]
                                     &       & 2.0       &\\
                 &        &          &       &           &3.6 \\
Magnet \#1       & 1.630  & 10.40/14.12\footnotemark[1]
                                     & 39.98 & 40.01     &\\
                 &        &          &       &           & 9.4\\
Magnet \#2       & 1.689  & 15.98/22.12\footnotemark[1]
                                     & 64.04 & 65.01     &\\
                 &        &          &       &           &9.4\\
Magnet \#3       & 1.628  & 28.04    & 94.00 & 70.01 &\\
                 &        &          &       &       & 429.7\\
Magnet \#4       & 1.583  & 30.04    & 94.02 & 38.01 &\\
                 &        &          &       &       & 101.0\\
Magnet \#5       & 1.607  & 30.06    & 94.00 & 55.01 &\\
                 &        &          &       &       & 15.0\\
Magnet \#6       & 1.611  & 30.02    & 94.04 & 55.00 & \\
                 &        &          &       &       & 300.0\\
                 &        &          &       &       & 337.0\\
Compr. tube      &        & 10.0     & 11.0  & 100.0 &\\
\end{tabular}
\end{ruledtabular}
\label{magnet-system}
\end{table}

The transmission $Tr$ of the system is defined as the fraction of tracks, ending within the
entrance of the feeding tube, to those passing the diaphragm in front of the first sextupole
magnet. For the four hyperfine states of hydrogen~\cite{HFS}, the calculations
yield $Tr(|1\rangle)\sim Tr(|2\rangle)=0.42$ (for both $\mu_{\rm eff}>0$), and
$Tr(|3\rangle)=0.001$, and $Tr(|4\rangle)=0.0004$ (for both $\mu_{\rm eff}<0$).

The performed tracking calculations do not account for intra-beam and residual-gas scattering. The
calculated transmissions thus only allowed one to estimate upper limits of the expected atomic
beam intensity $I_{\rm in}$ into the feeding tube. For a primary molecular flow $q({\rm H_2})$,
the intensity $I_{\rm in}(\rm{H})$ with atoms mainly in the states $|1\rangle$ and $|2\rangle$
($\mu_{\rm eff}>0$) was expected as
\begin{equation}
I_{in}({\rm H})=q({\rm H_2})\cdot 2\alpha\cdot\frac{\Omega}{2\pi}
    \cdot\frac{1}{4}\sum_{i=1}^{i=4}Tr(|i\rangle).
\label{I_est}
\end{equation}
For the degree of dissociation $\alpha$ a routine value of 0.8 (see e.g.,
Ref.~\cite{Wise_et_al_1993}) was assumed. $\Omega=0.022\pi$ is the solid angle covered by the
collimator aperture. The factor 1/4 reflects the assumption that the four substates in the atomic
beam from the nozzle are equally populated. For $q({\rm H_2})=1$\,mbar $l$/s or
$2.7\cdot 10^{19}{\rm H_2}$ molecules/s one expects $I_{in}(\rm H)\sim 1\cdot10^{17}$\,H atoms/s.

As described in the subsequent section, the rf  transition units are used to change the relative
occupation numbers of the states. The trajectory code allows one to simulate this change by
assigning a $\mu_{\rm eff}$ of one of the states to the atoms before they pass a magnet. As an
example, the medium-field transition unit (MFT) behind magnet No.~3 (see Fig.~\ref{ABS}) brings H
atoms from state $|2\rangle$ into state $|3\rangle$. This is simulated by assigning $\mu_{\rm
eff}(|2\rangle)>0$ to the atoms in the magnets $1-3$ and $\mu_{\rm eff}(|3\rangle)<0$ in the magnets
$4-6$, where they get deflected from the beam axis. This results in a small value
$Tr(|2\rangle)=0.017$. From this value and the above value $Tr(|1\rangle)=0.42$ the  vector
polarization is expected as
\begin{equation}
p_z = \frac{Tr(|1\rangle)-Tr(|2\rangle)}{Tr(|1\rangle)+Tr(|2\rangle)} = 0.91
\label{+PzHest}
\end{equation}
under the assumption of 100 \% efficiency of the transition unit.

The design and the properties of the permanent sextupole magnets~\cite{VACODYM} were discussed in an earlier
paper~\cite{Vassiliev_et_al_2000}. To achieve the pole-tip field values of $\sim$1.5\,T, each
magnet was produced from 24 segments employing three different types of NdFeB compounds. The
expected pole-tip values (Table~\ref{magnet-system}) and the precise radial dependence $B(r)\sim
r^2$ within the magnet apertures were confirmed. For the first time the predicted high multipole
components~\cite{Halbach_1980} up to a 102-pole structure very near to the aperture surface could
be measured~\cite{Vassiliev_et_al_2000}.

After the field measurements, the magnets were encapsulated to prevent diffusion of hydrogen into
the magnet material, which
might deteriorate the magnetic properties, and to avoid the pumping of gas from the sintered
magnet bodies. The housings were made from thin stainless steel cans of 0.2\,mm thickness for the
conical and cylindrical walls within the magnet apertures and 0.3\,mm for the front and end
covers. During the final welding to close the housings with magnets installed, the local
temperature of the magnet material had to be kept below the Curie temperature of
60\,$^{\circ}$C. This was achieved by welding with the use of a pulsed 15\,Hz Nd:YAG laser,
delivering 1.1\,J in a pulse of 2\,ms~\cite{FIL}.
Overlapping weld spots of $\sim$0.3\,mm diameter, set around the adjacent circular, 0.2\,mm thick
weld lips, allowed one to finish the housings with leak rates
$\sim10^{-10}\,\rm{mbar}\,\it{l}$/s. Inside the housings the magnets were fixed to suppress axial
and rotational movements caused by the force of the adjacent magnets. Finally the free slits
within the housings were filled by $\sim$20\,mbar krypton to enable leak tests by mass spectroscopy.
\subsection{Radio Frequency Transition Units}
The ABS is equipped with three types of transition units, a weak field, a medium field, and a
strong field rf transition unit (WFT, MFT, and SFT units). Together with the selecting properties
of the sextupole magnets, they enable one to achieve all vector and tensor polarizations of the
atomic hydrogen and deuterium gas in the storage cell. In all three units,
transitions between the hyperfine states, split according to the Breit-Rabi diagram  by a static
magnetic field (see e.g., Ref.~\cite{Haeberli_1967}), are induced by the magnetic component
($B_{\rm rf}$) of an rf field, leading to changes in the population of the states. The
static field $B_{\rm stat}$ consists of two parallel components, a homogeneous field $B_{\rm hom}$
and a superimposed weaker gradient field $B_{\rm grad}$, both orthogonal to the beam direction.
The field gradient along the beam direction is required to satisfy the condition of adiabatic
passage~\cite{Haeberli_1967,Abragam+Winter_1958}.

\begin{figure}[b]
\includegraphics[width=\columnwidth, angle=-90]{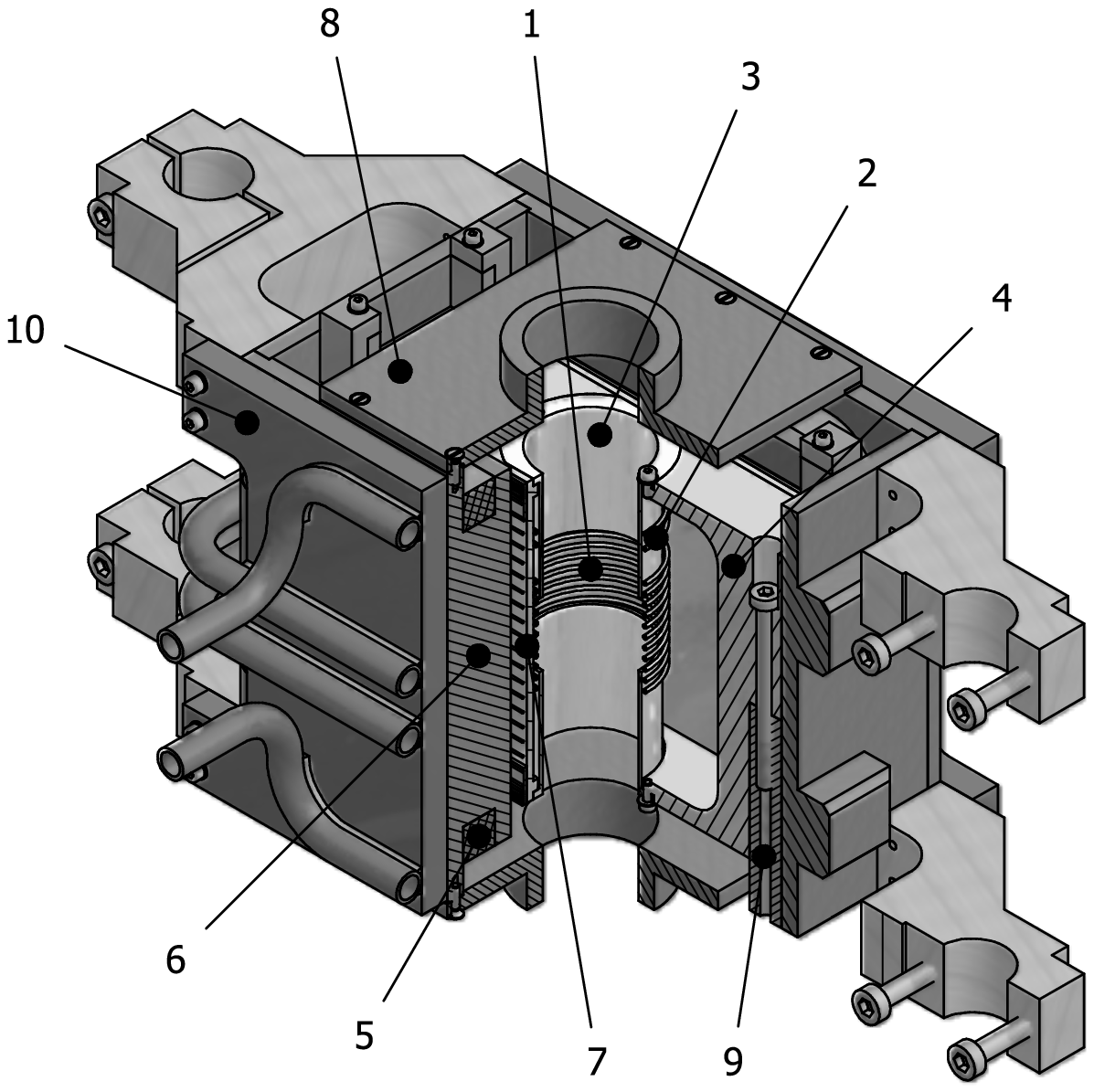}
\caption{Three-quarter-section view of the MFT unit with the support structure (1: self-supporting
rf coil with spacers, 2: pick-up loop, 3: Al tubes defining the length of the transition-inducing
rf field, 4: Cu cavity, 5: coil around the pole shoe (6), providing the static field $B_{\rm
stat}$, 7: slit between pole shoe and cavity wall housing the gradient-field coil, 8: components
of the static magnet yoke, also serving as shielding against external fields,
9: cavity-positioning element, 10: Cu pads cooled by means of water-carrying tubes. The cavity
with the rf coil and the pick-up loop can be taken out from the surrounding components.}
\label{MFT}
\end{figure}

\begin{figure}[hbt]
\includegraphics[width=\columnwidth]{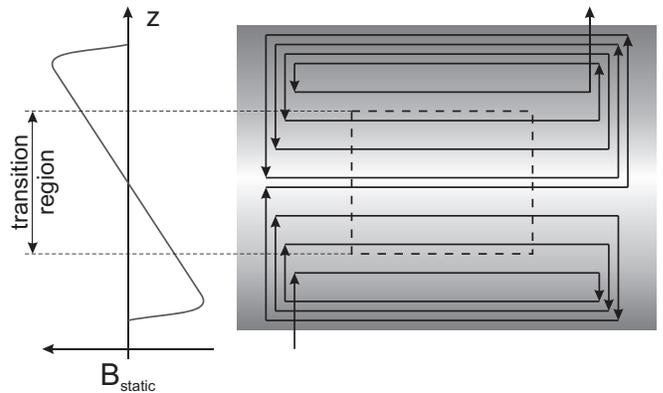}
\caption{Arrangement of the windings producing the static gradient field $B_{\rm grad}$ shown in
the left-hand side of the figure. In all transition units the field lies in the direction of the
static homogeneous field, the field gradient $dB/dz$ lies in the beam direction which defines the
$z$ axis. In $z$ direction,  the transition reagion (indicated by the dashed lines) is confined to
the range of constant gradient by the Al tubes, in orthogonal direction by the beam diameter.}
\label{GF}
\end{figure}

The assemblies of the WFT and the MFT units are similar~\cite{Lorenz_Dipl_1999}. The layouts
follow those of the units developed for the HERMES experiment~\cite{Gaul+Steffens_1992}. In both
units the rf field is produced by a coil with the axis along the beam direction, and consequently
$B_{\rm rf}$ orthogonal to $B_{\rm stat}$. The MFT unit is shown in Fig.~\ref{MFT}.
Figure~\ref{GF} schematically shows one of the grooved aluminum frames with the windings producing
the gradient field.
A WFT unit is operated in a weak magnetic field, $B_{\rm stat}\le$10\,G for hydrogen and $\le$5\,G
for deuterium, where the total atomic spin $F$ is a good quantum number. In hydrogen the $F=1$
levels $|1\rangle$, $|2\rangle$, and $|3\rangle$ with magnetic quantum numbers $m_{\rm F}=+1,\,0$,
and $-1$, respectively, can be regarded as equally spaced. In deuterium the same holds for the four
$F=3/2$ levels $|1\rangle$, $|2\rangle$, $|3\rangle$, and $|4\rangle$) with  $m_{\rm
F}=+3/2,\,+1/2,\, -1/2$, and $-3/2$, respectively,  and for the two $F=1/2$ levels $|5\rangle$ and
$|6\rangle$ with $m_{\rm F}=-1/2$ and $+1/2$, respectively. The magnetic component of the rf dipole
field induces transitions between each pair of neighboring $m_{\rm F}$ states with $\Delta m_{\rm
F}=\pm 1$. $|\Delta m_{\rm F}|=2$ transitions are forbidden. The interchange of the population
between the states $|1\rangle$ and $|3\rangle$ in hydrogen, e.g., is caused by a two-quantum
transition via the intermediate state $|2\rangle$. In the classical description of the adiabatic
passage method~\cite{Abragam+Winter_1958} the population change should not depend on the sign of
the magnetic field gradient relative to the beam direction. An exact quantum-mechanical
treatment~\cite{Oh_1970,Schieck_2008}, however, indicates that cleaner population changes from
state $|1\rangle$ to $|3\rangle$ in hydrogen and from state $|1\rangle$  to $|4\rangle$ in
deuterium are obtained with a negative field gradient, i.e., a $B_{\rm rf}$ field decreasing in
the beam direction. Deviations from adiabaticity are discussed in
Ref.~\cite{Oh_1970,Philpott_1987}.

The MFT unit is operated at higher values of $B_{\rm stat}$, where the differences in the energy
spacings of pairs of hyperfine states with $\Delta m_{\rm F}=\pm 1$ allow one to select single
transitions. Originally developed for an polarized alkali ion source~\cite{Jaensch_et_al_1985},
the MFT unit now is a standard component in polarized hydrogen and deuterium sources as discussed,
e.g., in Ref.~\cite{Roberts_et_al_1992}. Appropriate choice of $B_{\rm hom}$, $B_{\rm grad}$, and
the rf frequency allows one to induce selected transitions $|1\rangle\leftrightarrow|2\rangle$ and
$|2\rangle\leftrightarrow |3\rangle$ in hydrogen or $|1\rangle\leftrightarrow |2\rangle$,
$|2\rangle\leftrightarrow |3\rangle$, and $|3\rangle\leftrightarrow |4\rangle$ in deuterium.
Furthermore, the choice of the field gradient allows one to achieve consecutive transitions. As an
example a negative field gradient in the MFT unit behind the first set of magnets, i.e., a $B$
field decreasing in beam direction, at a fixed rf frequency leads to the sequence of the
transitions $|3\rangle\rightarrow|4\rangle$, $|2\rangle\rightarrow |3\rangle$, and finally
$|1\rangle\rightarrow|2\rangle$ in deuterium, leaving the state $|1\rangle$ empty.

\begin{figure}[t]
\includegraphics[width=5cm, angle=-90]{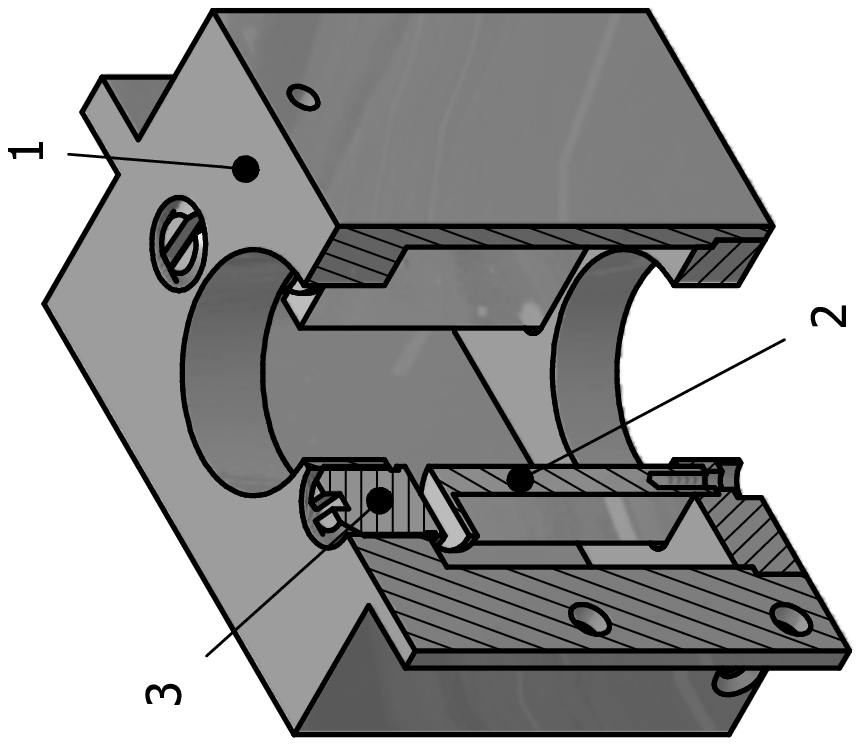}
\caption{Three-quarter-section view of the rf cavity of the SFT unit for deuterium (1: the two
resonant-field creating conductors, 2: the adjustible capacitor plates, 3: Cu cavity). The inner
dimensions of the cavity are 56\,mm along $B_{\rm stat}$, 36\,mm orthogonal to it and 36\,mm
height. The cross section of the conductors is $14\times4\,{\rm mm}^2$.}
\label{SFT}
\end{figure}

The SFT unit is used to induce transitions between states in the upper and lower hyperfine
multiplet in hydrogen and deuterium. Contrary to the historical name, indicating a strong magnetic
field, the SFT unit is operated with magnetic fields comparable to those used in the MFT unit. The
transition frequencies are comparable with those of the hyperfine splitting (1420\,MHz for
hydrogen and 327\,MHz for deuterium), and thus are much higher than those in the WFT and MFT
units. The rf field in a SFT unit is produced by a twin-line resonator inside a Cu box tuned to
the $\lambda/4$ resonance~\cite{Capiluppi_2012}. The SFT unit~\cite{PNPI} is shown in Fig.~\ref{SFT}. Again, the layout follows that of the
unit used in the HERMES experiment~\cite{Gaul+Steffens_1992}. Two variable capacitors at the free
ends of the conducting rods, fed by the rf power with a relative phase shift of 180\,$^{\circ}$,
allow one to tune the device.
\subsection{Slow Control System}
Industrial components, providing reliable and long-term support, were selected for the control
system of the whole setup consisting of the ABS and the diagnostics tools, the storage cell
positioning system, the Lamb-shift polarimeter, and the supply system of a calibrated flow of
unpolarized molecular gas. The interlock system has been implemented on the basis of SIEMENS
SIMATIC S7-300 family of programmable logic controllers. In order to unify the interfacing to
the control computer, all front-end equipment is connected via the PROFIBUS DP fieldbus. The
process control software was implemented using the Windows-based WinCC toolkit from SIEMENS.
The system controls the operation of the pumps and the valves. It reads the pressure gauges
and controls the regeneration cycles of the cryopumps. Via a control network, the temperature
of the nozzle is stabilized within $\pm 0.5$\,K. Furthermore, all power-supply units, rf
generators and amplifiers are set and controlled. The whole variety of components to be
controlled, the logical structure of the control and interlock system, and a separate device
for parameter studies are described in Ref.~\cite{Kleines_et_al_2006}.

\section{Studies of the free hydrogen jet\label{Sec:III}}
\subsection{Atomic beam profile near the nozzle}
\begin{figure}[t]
\includegraphics[width=7cm, angle=-90]{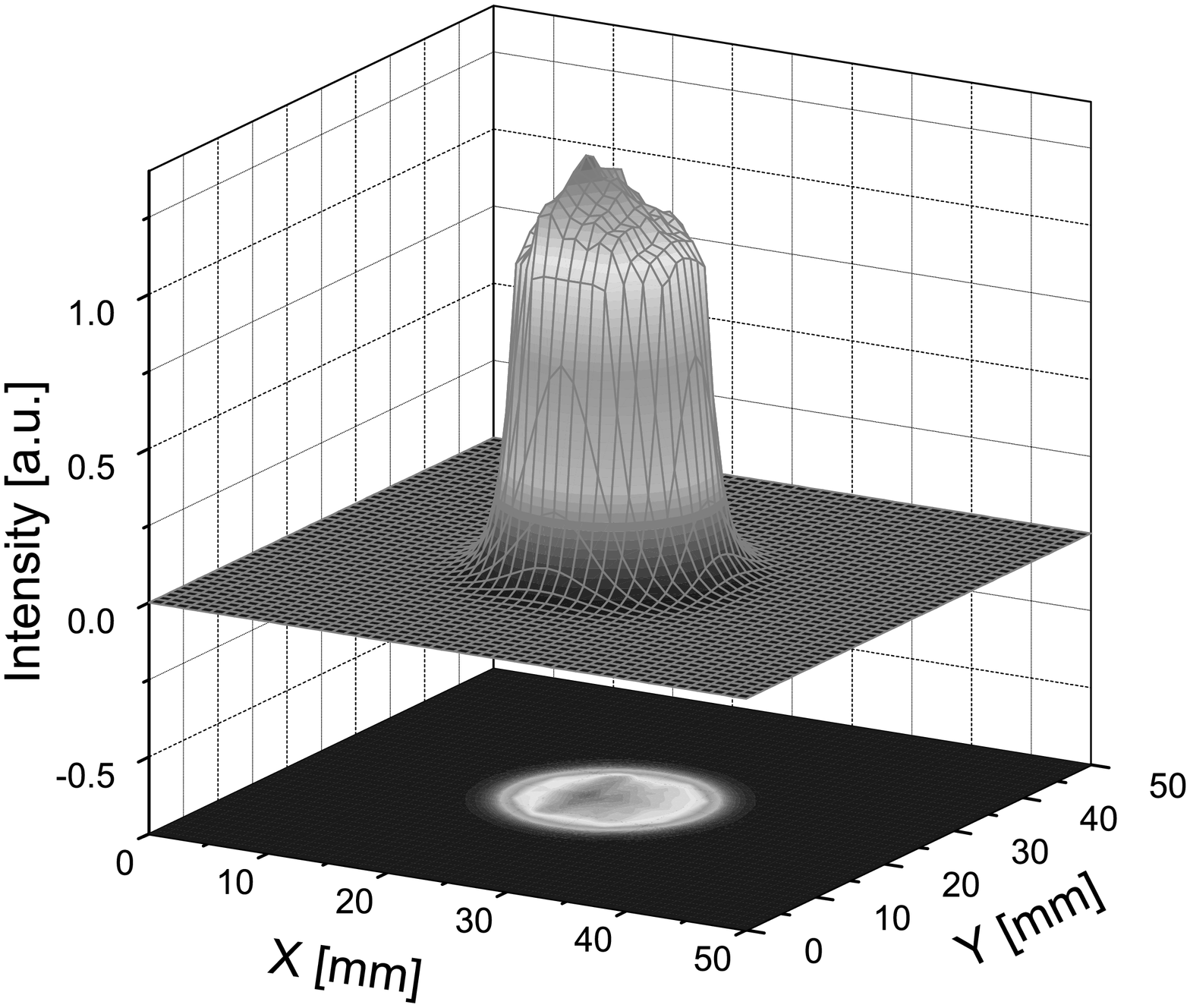}
\caption{Two-dimensional profile of the atomic hydrogen beam 10\,mm from the nozzle, deduced from
recombination heating of gold-plated tungsten wires of 5\,$\mu$m in a $8\times8$ wire grid.}
\label{MWM}
\end{figure}

A novel device has been used to measure the profile of an atomic beam via the deposition of
recombination heat on thin wires in a two-dimensional
grid~\cite{Vassiliev_et_al_1998,Vassiliev_et_al_1999}.
Atoms stuck on the surface of gold-plated tungsten wires of 5\,$\mu$m diameter recombine and
are reemitted as molecules. The recombination heat (4.476\,eV per hydrogen molecule) leads to
a change of temperature and, thus, resistance along each wire. The measurement of the resistance
changes of all the wires in the grid allows one to deduce the center and the profile of the beam.
Figure~\ref{MWM} shows the beam profile resulting with a $8\times8$ wire grid positioned between
skimmer and collimator, performed as a first proof of the method.
Later, such a device has been used to compare measured and calculated beam profiles along the beam
axis between nozzle and skimmer~\cite{Nass+Steffens_2009}.
\subsection{Degree of dissociation of the free atomic jet}
The dissociation of the primary molecules is achieved by the interaction of the electrons and the
hydrogen or deuterium molecules in the plasma of the dissociator. The degree of dissociation of
the beam from the nozzle depends on the rf power, applied to maintain the plasma, the primary
molecular gas flow into the dissociator, and the temperature of the nozzle and the lower end of
the discharge tube. These dependencies have been studied before installation of the sextupole
magnets with a setup containing a crossed-beam quadrupole mass
spectrometer~\cite{Max_Dipl_1999,Max_et_al_1999}.

\begin{equation}
\alpha=\frac{\rho_{\rm a}}{\rho_{\rm a}+2\cdot \rho_{\rm m}},
\label{alpha-definition}
\end{equation}
The admixture of molecules in an atomic beam is described by the degree of dissociation,
where $\rho_{\rm a}$ and $\rho_{\rm m}$ are the densities of atomic and molecular hydrogen or
deuterium in the beam. Other authors (e.g., Ref.~\cite{Nass_et_al_2003}) use the atomic and
molecular intensities $I_{\rm a}$ and $I_{\rm m}$ in the definition of the degree of dissociation
($\alpha_{\rm I}$) in Eq.~(\ref{alpha-definition}). The two definitions of are related by
\begin{equation}
\frac{I_{\rm m}}{I_{\rm a}}=\frac{\bar{v}_{\rm m}}{\bar{v}_{\rm a}}\cdot\frac{1-\alpha}{2\alpha}=
\frac{1-\alpha_{\rm I}}{2\alpha_{\rm I}}\,.
\label{alpha-definition1}
\end{equation}

\begin{figure}[hbt]
\includegraphics[width=\columnwidth]{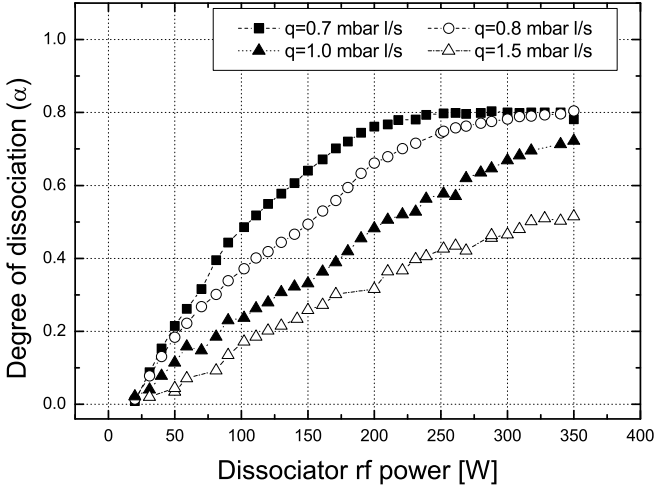}
\caption{Degree of dissociation $\alpha$ of the free hydrogen jet as function of the applied rf
power for different primary
molecular hydrogen flows and a nozzle temperature of 70\,K.}
\label{AlphaRF}
\end{figure}

This quantity was determined with the quadrupole mass spectrometer (QMS) in
a conventional way as
\begin{eqnarray}
\alpha&=&\frac{S^{*}_{\rm a}}{S^{*}_{\rm a}+2\,k_{v}\,k_{\rm ion}\,k_{det}S_{\rm m}}.
\label{alpha free jet}
\end{eqnarray}
Here, $S^{*}_{\rm a}=S_{\rm a}-\delta S_{\rm m}$ denotes the atomic signal, corrected for
dissociative ionization. The parameter $\delta = 0.0141$ was obtained following the method
described in Ref.~\cite{Koch+Steffens_1999}. The coefficient
$k_{v}=\overline{v}_{m}/\overline{v}_{a}$, accounting for the difference in atom and molecule
velocity, was chosen as $1/\sqrt{2}$ under the assumption of thermalization of the beam emerging
from the nozzle. Furthermore, $k_{{\rm ion}}=0.64$ ~\cite{ionization CS} accounts for the
differences in ionization cross section for atomic and molecular hydrogen, and
$k_{{\rm det}}=0.84$ for the detection probability~\cite{Max_Dipl_1999}. As an example of the
parameter studies, Fig.~\ref{AlphaRF} shows the deduced dependencies on the rf power for a set
of primary molecular hydrogen gas flows. For typical flow values
$q({\rm{H_2}})\le1.0\,{\rm mbar\,\it{l}/s}$ a saturation value around 0.8 was obtained.

\section{Beam intensity\label{Sec:IV}}
\begin{figure}[t]
\includegraphics[width=7.5cm]{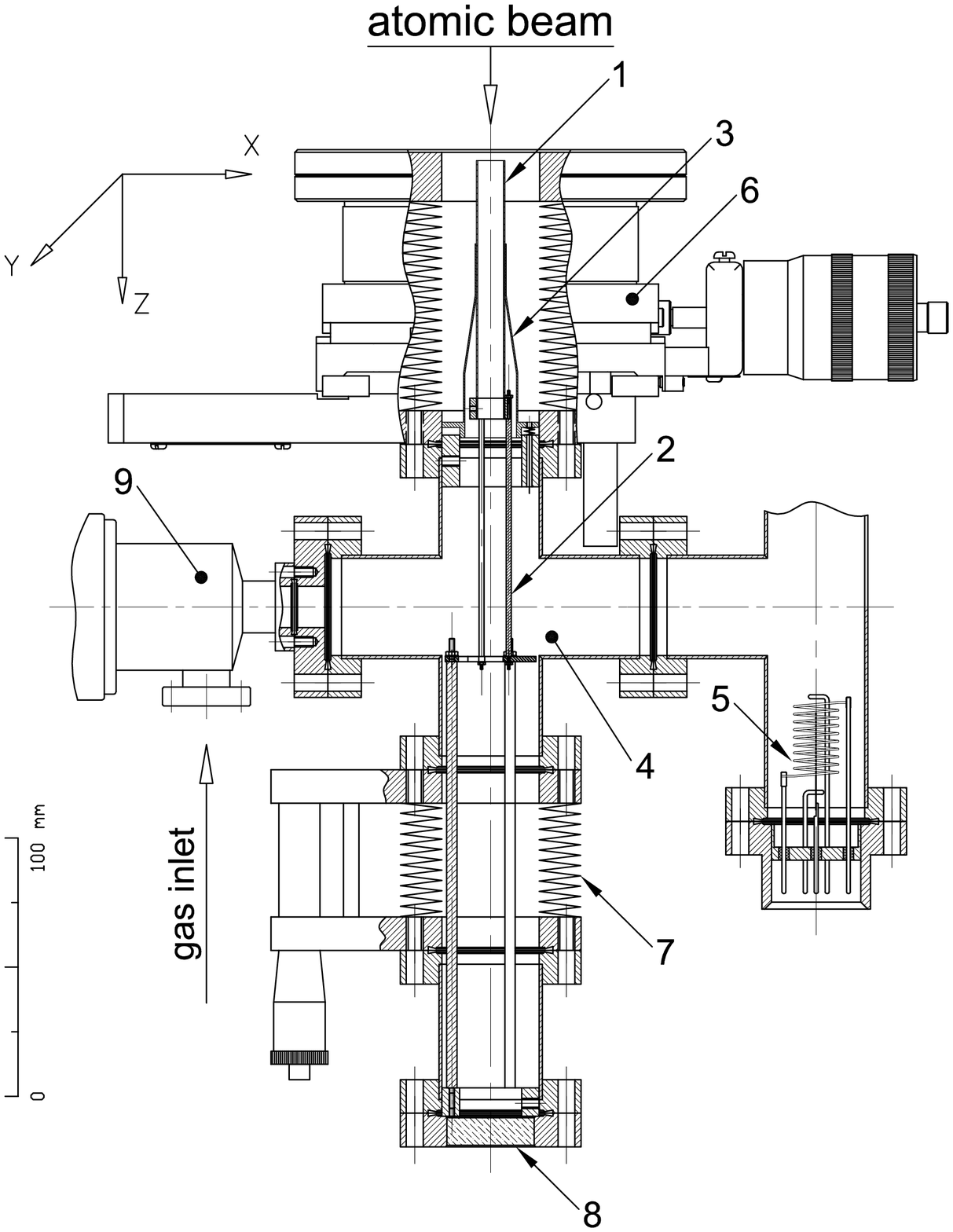}
\caption{Side view of the compression-tube setup, made from standard ultra-high-vacuum components,
with a partial cut along the axis.  (1: compression tube, 2: support of the compression tube based
on the lower flange,  3: narrow tube around 1 closing the upper volume and allowing axial shifts
of the tube by the support, 4: compression volume, 5: hot-cathode pressure gauge, 6: $xy$
manipulator, 7: $z$ manipulator, 8: glass viewport, 9: electromagnetic valve for gas inlet.}
\label{CTSetup}
\end{figure}

The intensity of the polarized beam from the ABS together with the layout of the storage cell
determines the areal density of the target gas. The intensity of the beam has been measured with
the use of a compression-tube setup~\cite{Nekipelov1,Nekipelov2}, shown in Fig.~\ref{CTSetup},
to optimize the ABS operation parameters. The measurements were performed at a 300\,mm distance
from the compression-tube entrance to the last magnet and an inner tube diameter of 10.0\,mm as
set in the tracking calculations. The length of the compression tube of 100\,mm was made equal to
that of the foreseen feeding tube of the storage cell. The narrow tube around the compression tube
on a support, based on the lower flange, separates the volume around the tube from the compression
volume. The $xy$ manipulator serves for centering the tubes and for intensity-profile
measurements. The construction allows axial shifts of the compression tube by the $z$ manipulator
and the use of tubes of different diameters.

\begin{figure}[t]
\begin{minipage}[t]{\columnwidth}
\includegraphics[width=6.5cm,angle=-90]{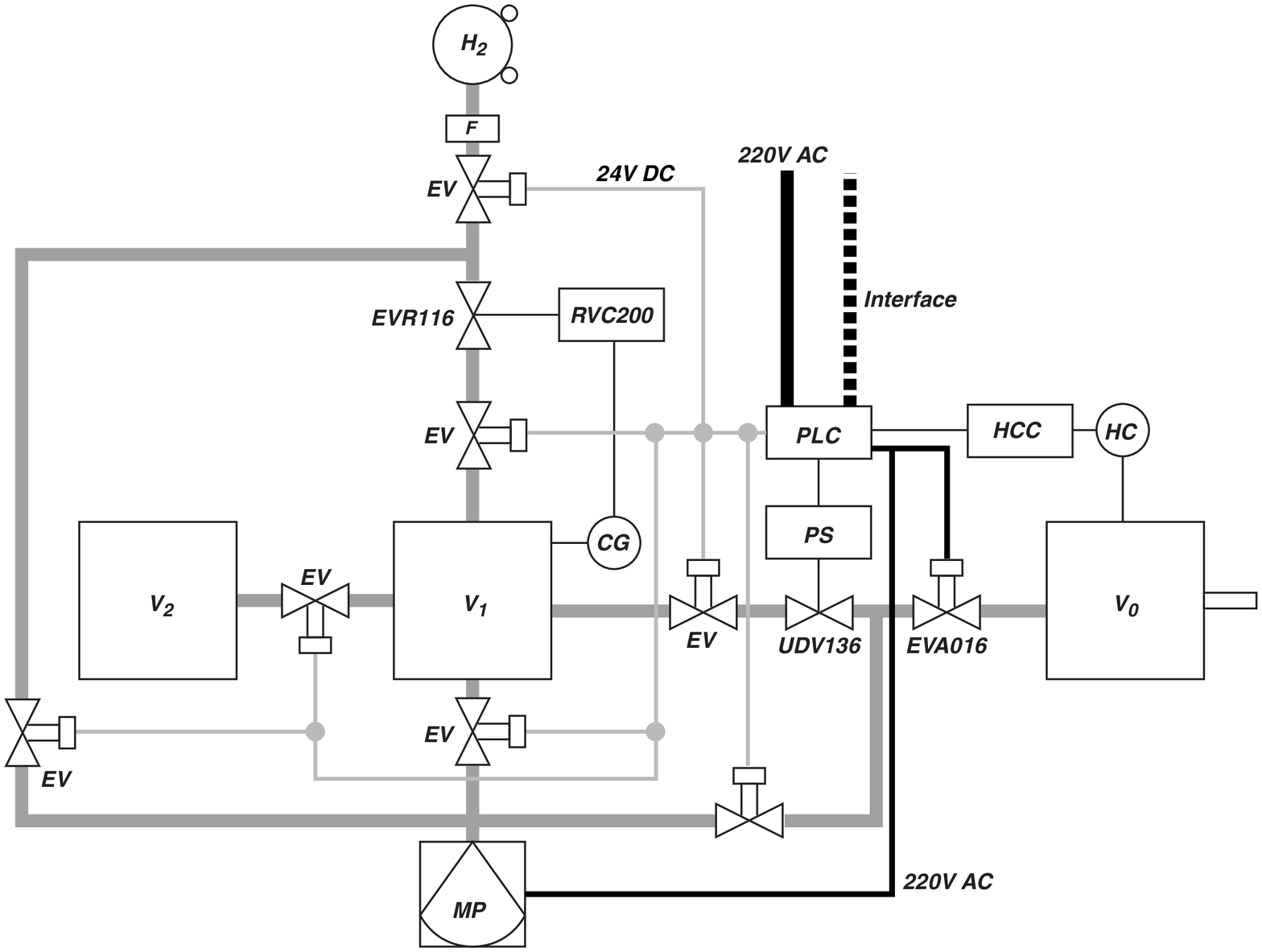}
\caption{Source of calibrated molecular gas flow (V$_{0}$: compression chamber, V$_{1}$:
gas-storage chamber feeding V$_{0}$ via the needle valve UDV136\protect\footnote{All the
valves and the gauge controller are supplied by Pfeiffer Vacuum GmbH, D--35614 Asslar, Germany
(manufacturer Balzers AG, Liechtenstein).}, V$_{2}$: chamber of calibrated
volume used to determine that of V$_{1}$). The pressure in V$_{1}$ is measured by the capacitance
gauge CG and is kept constant by the dosing valve EVR116 with the gauge controller RVC200. The
whole setup, including the evacuation elements can be operated manually or by the programmable
logic controller PLC either within the ABS control system~\cite{Kleines_et_al_2006} or as a
separate system.}
\label{UGSS}
\end{minipage}
\end{figure}
\begin{figure}[hbt]
\includegraphics[width=\columnwidth, angle=0]{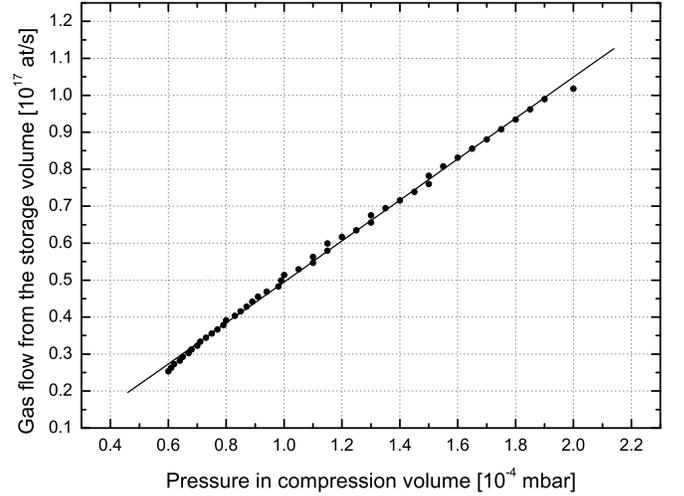}
\caption{Calibration curve for hydrogen used to deduce from the measured pressures the intensities
of the hydrogen and deuterium beam injected into the compression tube.}
\label{Calibration}
\end{figure}

The intensity of the beam, entering the compression volume through the compression tube, is
measured via the pressure in the compression volume. It is determined by the equilibrium between
the incoming beam intensity $I_{\rm in}$ and the outgoing intensity $I_{\rm out}$. Under the
assumption of a pure atomic beam and complete recombination in the compression volume
\begin{eqnarray}
I_{\rm in}({\rm atoms/s})&=&2\cdot I_{\rm out}({\rm molecules/s})\nonumber\\
 &=&2\cdot\Delta P \cdot C_{\rm tube}\nonumber\\
 &=&2\cdot\Delta P \cdot 1.03 \cdot 10^{20}\cdot\frac{d^{3}}{l}\sqrt{\frac{T}{M}}~.
\label{I_in-DeltaP}
\end{eqnarray}
Here $\Delta P$ is the difference between the pressure measured in the compreesion volume and that
in the ABS chamber V. The conductance of the compression tube, $C_{\rm tube}$, is
determined by the inner diameter $d$ of the tube, its length $l$, the gas temperature $T$, and the
molar mass $M$ of the gas (given in cm and K, respectively)~\cite{Roth_Vakuum}. The factor
2 takes into account that the same pressure is measured in the hot-cathode gauge for
$2\cdot I_{\rm in}$\,(H atoms/s) and $1\cdot I_{\rm in}$ (H$_2$ molecules/s). For $d=10$\,mm,
$l=100$\,mm, $T=290$\,K, and $M=2$ for hydrogen pressure differences $\Delta P$ on the order of
$10^{-4}$\,mbar are expected for atomic hydrogen beam intensities in the order of
$10^{17}$\,atoms/s. The relation between $I_{\rm in}$ and $\Delta P$ for hydrogen has been
determined experimentally with the use of a source of calibrated molecular hydrogen gas
flow~\cite{Nekipelov1, Nekipelov2}, depicted in Fig.~\ref{UGSS}. The measured dependence with a
linear fit is shown in Fig. \ref{Calibration}. The calibration curve allows one to determine
absolute values of $I_{\rm in}$ of hydrogen and deuterium beams. The calibration for deuterium was
deduced from the one for hydrogen by scaling with a factor $1/\sqrt{2}$, according to
Eq.~(\ref{I_in-DeltaP}).

\begin{figure}[t]
\includegraphics[width=\columnwidth, angle=0]{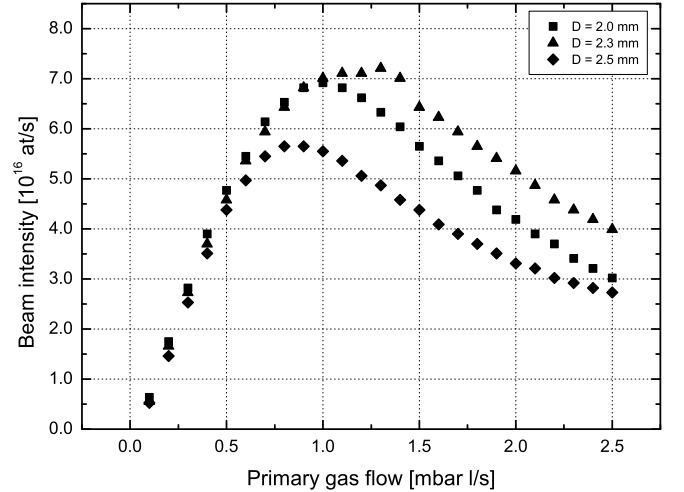}
\caption{Intensity of the hydrogen beam (states $|1\rangle$ and $|2\rangle$) injected into the
compression tube as function of the primary molecular gas flow for different nozzle diameters $D$
(nozzle temperature 60\,K, dissociator power 300\,W).}
\label{I(gasflow)}
\end{figure}
\begin{figure}[h]
\includegraphics[width=\columnwidth, angle=0]{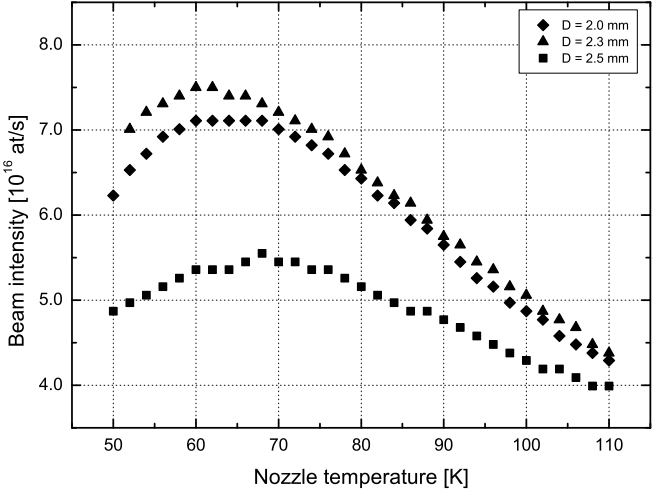}
\caption{Intensity of the hydrogen beam (states $|1\rangle$ and $|2\rangle$) injected into the
compression tube as function of the nozzle temperature for different nozzle diameters $D$ (primary
molecular gas flow 1\,mbar\,$l$/s, dissociator power 300\,W).}
\label{I(temperature)}
\end{figure}
\begin{figure}[h]
\includegraphics[width=\columnwidth, angle=0]{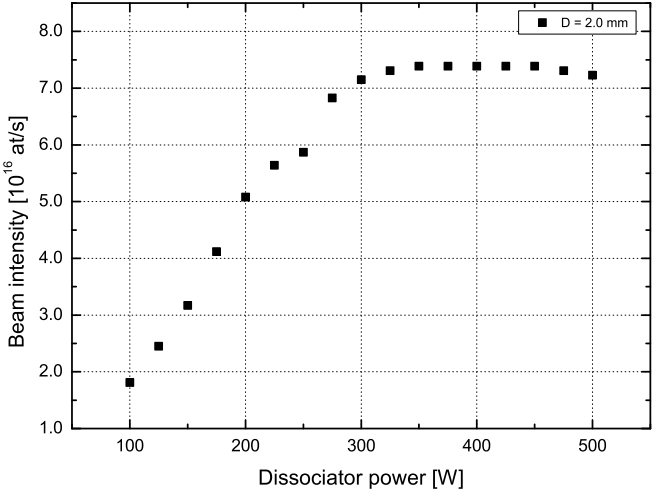}
\caption{Intensity of the hydrogen beam (states $|1\rangle$ and $|2\rangle$) into the compression
tube as function of the dissociator power for a nozzle diameter of 2 mm (nozzle temperature
$60$\,K, primary molecular gas flow 1\,mbar\,$l$/s).}
\label{I(dissopower)}
\end{figure}
The dependences of $I_{\rm in}$ on the dissociator-operation parameters primary molecular hydrogen
flow $q({\rm H_2})$,  nozzle temperature $T_{\rm n}$, and dissociator power $P_{\rm diss}$ have
been studied to find the optimum values. They are shown in the
Figs.~\ref{I(gasflow)},~\ref{I(temperature)}, and~\ref{I(dissopower)}, respectively, for different
nozzle-orifice diameters. The figures show that for the hydrogen beam (states $|1\rangle$ and
$|2\rangle$) with the standard operation parameters $q_{\rm H_2}=1.1$\, mbar $l$/s,
$T_{\rm n}=70$\,K, $P_{\rm diss}=350$\,W, and with a nozzle-orifice diameter of 2.3\,mm an
intensity of $I_{\rm in}({\rm H})=(7.5\pm 0.2)\cdot10^{16}$ particles/s is achieved, quite close
to the earlier estimate from Eq.~(\ref{I_est}). Besides the dominant atomic component of H atoms,
this value includes  small admixtures of H atoms in state $|3\rangle$ and molecular hydrogen. The
first kind can be estimated with the use of the calculated transmissions
(Sec.~\ref{Sec:II-Magnets}) as $0.017/0.84\approx2\%$. The amount of the second admixture has
been measured as described below.

For the deuterium beam  (states $|1\rangle$, $|2\rangle$, and $|3\rangle$) the optimization
procedure gave an intensity of
$I_{\rm in}({\rm D})=(3.9\pm 0.2)\cdot10^{16}$\,particles/s, achieved with $q({\rm D_2})=0.9$\,
mbar $l$/s, $T_{\rm n}=65$\,K, and $P_{\rm diss}=300$\,W, slightly lower than those for hydrogen.

\section{Hydrogen beam profiles\label{Sec:V}}
Beam profiles were measured at various positions at various positions behind the last sextupole
magnet with the use of\\
$\hspace*{5mm}\bullet$ a compression tube of reduced dimensions (5\,mm diameter)\\
$\hspace*{5mm}\bullet$ a crossed-beam quadrupole mass spectrometer, and\\
$\hspace*{5mm}\bullet$ a supplementary method of reduction of MoO$_{3}$ by hydrogen.
\subsection{Measurements with the compression tube\label{Sec:V-CTprofile}}
\begin{figure}[hbt]
\includegraphics[width=\columnwidth]{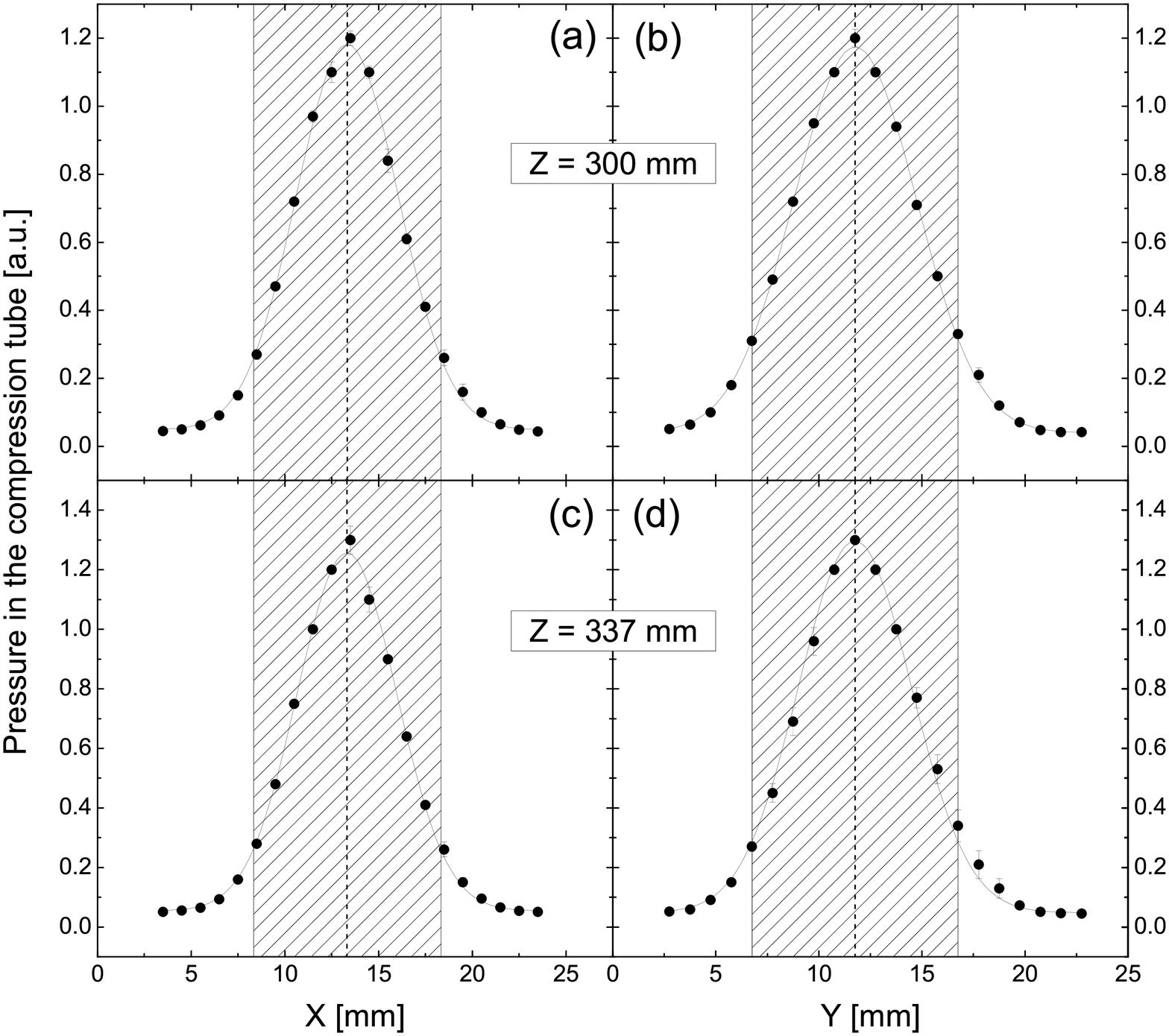}
\caption{Cross sections of the beam profile in the mid-plane, measured with compression tube of
5\,mm diameter and 50\,mm length. Measurements in the $xz$-plane (a, c) and $yz$-plane (b, d)
performed at two different positions: $z=300$\,mm (a, b) and $z=337$\,mm (c, d) behind the last
sextupole magnet of the ABS. The shaded area represents position and dimensions of the compression
tube used in intensity measurements.}
\label{CT-xy}
\end{figure}

For the determination of the beam dimensions at two positions, 300\,mm and 337\,mm behind the
last magnet, the compression tube setup (Fig.~\ref{CTSetup}) was used, making use of the
possibility of axial movement by the $z$ manipulator and of that to install a narrower and shorter
compression tube of 5\,mm diameter and 50\,mm length to enhance the spatial resolution. The $xy$
manipulator provided a lateral displacement of the compression tube by $\pm10$ mm in $x$ and $y$
direction. The center coordinates of the geometrical axis of the source had been determined
with the use of a bi-directional laser, centered inside the bore of the central support plate
(see Fig. \ref{ABS}). The relative intensity distributions in the $xz$ and $yz$ planes, given
by the measured pressure in the compression volume, are shown in Fig.~\ref{CT-xy}. Fits by
Gaussian distributions to the data yield full widths at half maximum
$\Gamma_{x}=(6.42\pm0.09)$\,mm,
$\Gamma_{y}=(6.99\pm0.06)$\,mm for the distributions measured at $z=300$\,mm and
$\Gamma_{x}=(6.27\pm0.08)$\,mm,
$\Gamma_{y}=(6.58\pm0.08)$\,mm at 337\,mm.

The center of gravity of the measured profile, defined as
\begin{equation}
r_{c}=\frac{\sum\limits_{i,j}\sqrt{x_i^2+y_j^2}\cdot P(x_{i},y_{j})}{\sum\limits_{i,j} P(x_{i},y_{j})},
\end{equation}
where $x_i$ and $y_j$ give the position of the compression-tube axis and $P(x_{i},y_{j})$ is the
pressure measured in the compression volume. The resulting $r_c$ shows a deviation of 0.12 mm from
the geometrical axis of the source. Furthermore, the data measured with the narrow compression
tube of 2.5\,mm radius can be used to derive the fraction of the beam entering the compression
tube of 5\,mm radius used in the intensity measurement of Sec.~\ref{Sec:IV}. The ratio
\begin{equation}
\eta=\frac{\sum\limits_{0}^{r_i\leq 2.5~{\rm mm}}P(x_{i},y_{j})}
                        {\sum\limits_{0}^{r_i\leq 10~{\rm mm}}{P(x_{i},y_{j})}},
\end{equation}
where $r_i$ is the distance of the compression-tube axis to the beam axis, yields
$\eta\approx0.7$.
\subsection{Measurements with the QMS\label{Sec:V-QMSProfile}}
\begin{figure}[b]
\includegraphics[width=6cm]{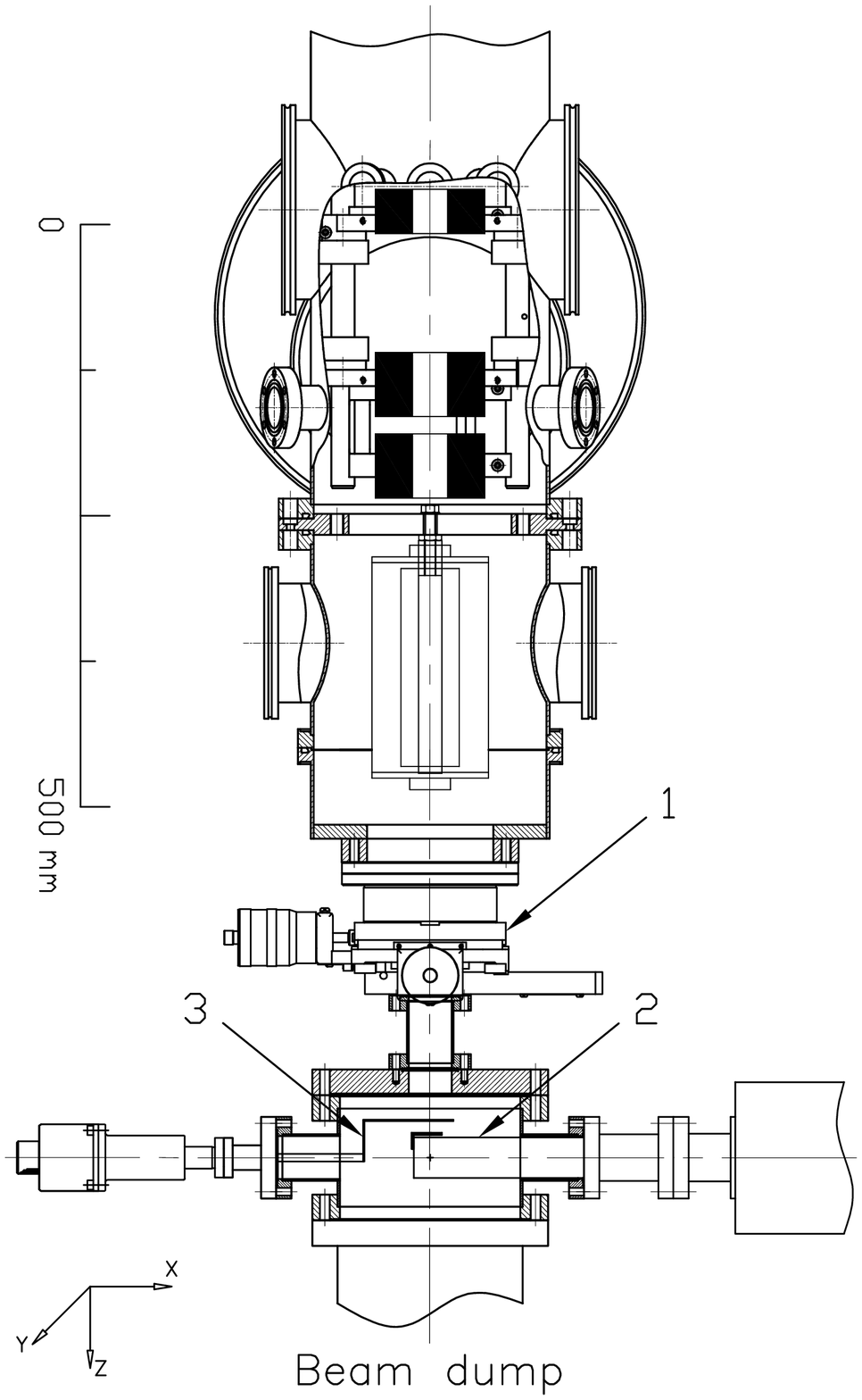}
\caption{Setup for the measurements of the beam profile with the QMS. (1): $xy$-table
enabling two-dimensional displacement of the entrance window of the QMS against the
geometrical axis of the ABS; (2): the QMS; (3): manually operated beam shutter. The
beam dump is an axially mounted cryo pump.}
\label{QMS-Setup}
\end{figure}
The beam-profile studies of Sec.~\ref{Sec:V-CTprofile} were extended with a setup utilizing a
crossed-beam quadrupole mass spectrometer (QMS) in the setup of Fig.~\ref{QMS-Setup}. Contrary to
the measurements with the compression tube, those with the QMS allow to separate the atomic and
molecular fractions in the beam. A 2\,mm diameter aperture was installed at the entrance of the
sensitive volume of the QMS to improve the resolution compared with that achieved by the
compression tube of 5\,mm diameter used in measurements of the preceding section. The layout
of the setup, presented in Fig.~\ref{QMS-Setup}, shows that in the present case the profile could
not be measured at a distance of $z=300$\,mm to the last magnet. Instead, measurements were
performed at $z=567$\,mm and, with installation of an extension tube, at $z=697$\,mm. The $xy$
manipulator enabled displacements of the aperture axis from the geometrical axis of the source in
any direction within limits set by the bore diameter of the $xy$ manipulator.

\begin{figure}[hbt]
\includegraphics[width=6.5cm, angle=-90]{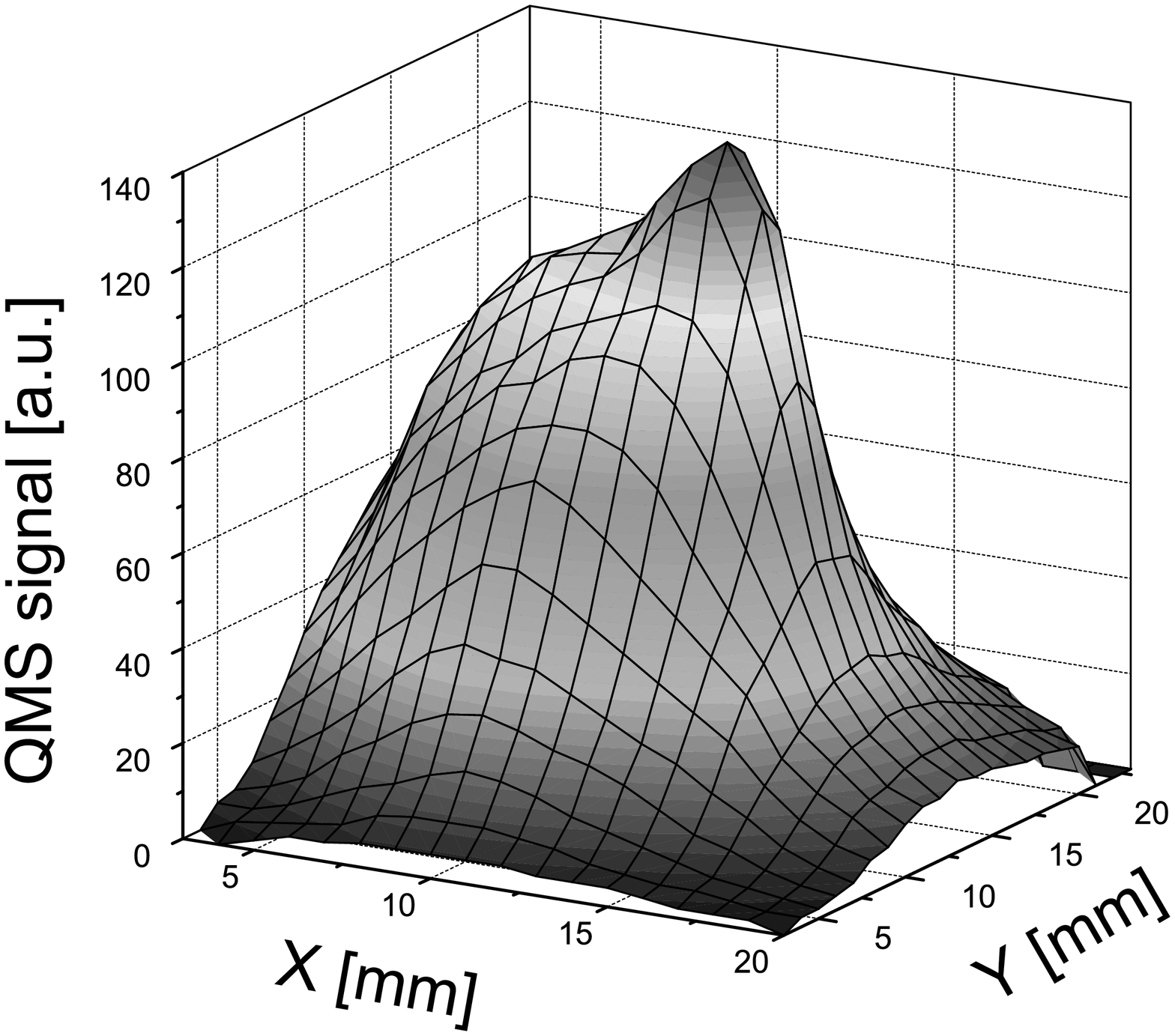}
\caption{Two-dimensional distribution of the atomic hydrogen component of the beam at $z=567$\,mm
before the nozzle-to-skimmer adjustment, showing a disinct deviation from azimuthal symmetry.}
\label{Profile1}
\end{figure}

\begin{figure}[t]
\begin{minipage}[t]{8cm}
\includegraphics[width=\columnwidth]{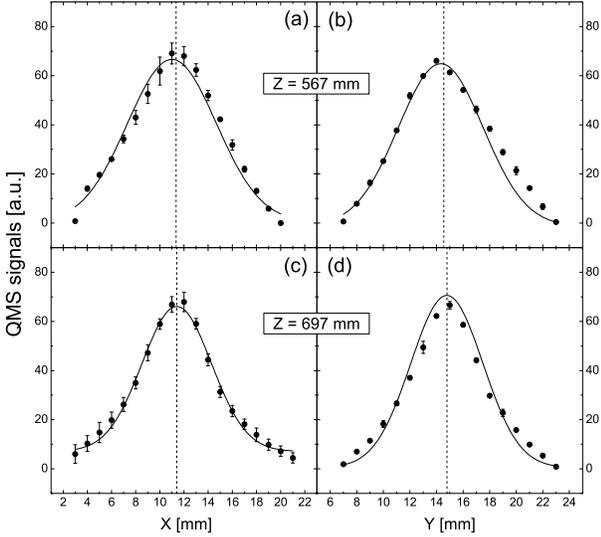}
\caption{Profiles of the atomic hydrogen component in the beam, measured with
the QMS 567\,mm and 697\,mm behind the last magnet.}
\label{QMS-xy}
\end{minipage}
\end{figure}

The first measured distribution of the atomic hydrogen (Fig.~\ref{Profile1}) showed a distinct
deviation from azimuthal symmetry, indicating an insufficient relative alignment of nozzle and
skimmer. The three threaded rods, supporting the dissociator with the nozzle via the
three-legged plate (label 2 in Fig.~\ref{ABS}), allow one to vary the position of the nozzle
relative to that of the skimmer while
\begin{figure}[hbt]
\includegraphics[width=6.5cm, angle=-90]{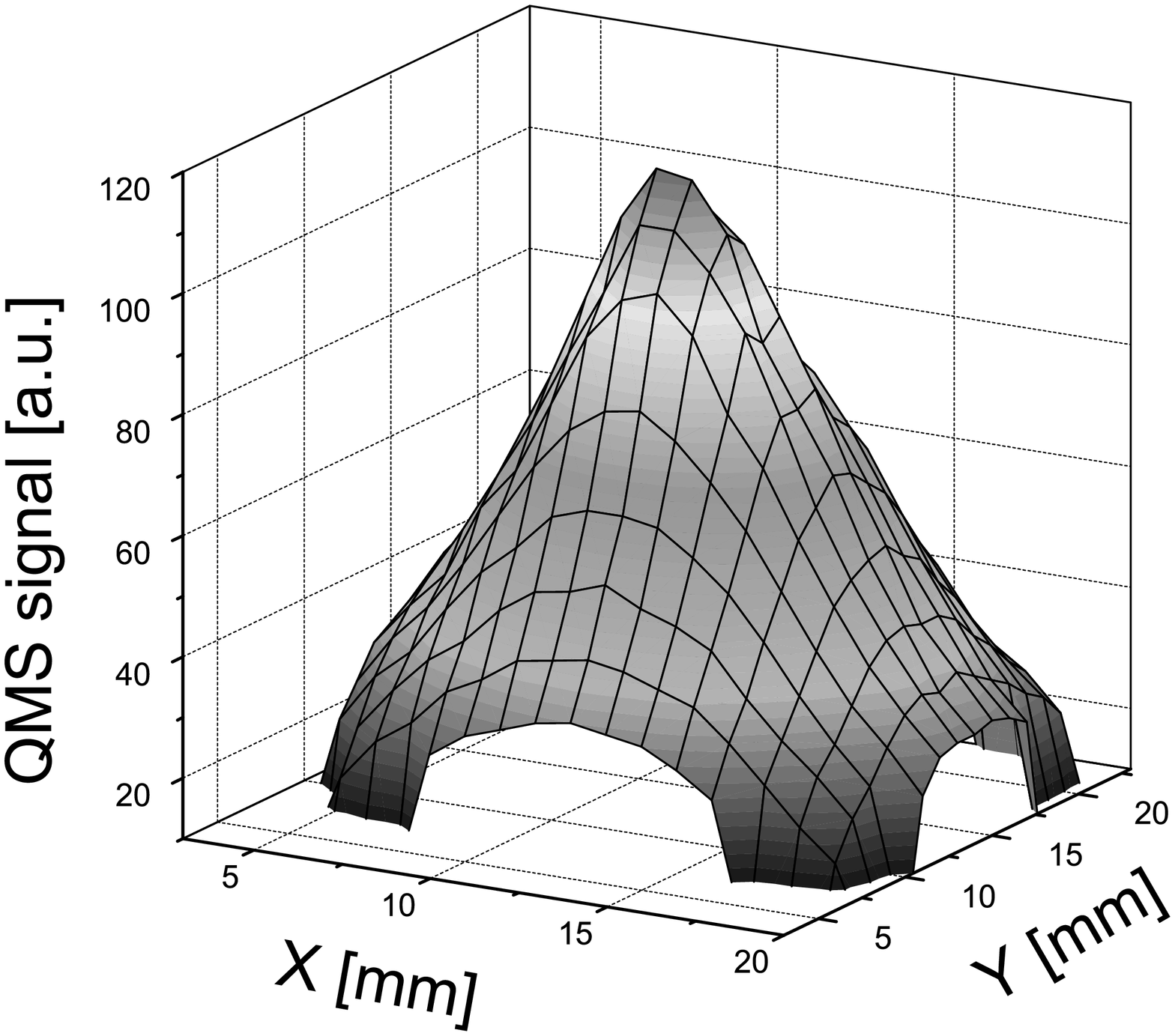}
\caption{The distribution corresponding to that of Fig.~\ref{Profile1} after nozzle-to-skimmer
adjustment resulting in azimuthal symmetry.}
\label{Profile2}
\end{figure}
the source is running. This possibility has been used to find a nozzle position which results
in an azimuthally symmetric distribution. The achieved symmetric distribution is shown in
Fig.~\ref{Profile2} and profiles of the atomic hadrogen component in the beam,
measured in $x$ and $y$ direction at $z=567$\,mm and $z=697$\,mm, are presented in Fig.~\ref{QMS-xy}.
Fits by Gaussian distributions to the data yield full widths at half maximum
$\Gamma_{x}=(7.36\pm0.43)$\,mm,
$\Gamma_{y}=(6.68\pm0.80)$\,mm for the distributions measured at $z=567$\,mm and
$\Gamma_{x}=(6.69\pm0.22)$\,mm, $\Gamma_{y}=(6.38\pm0.27)$\,mm at 697\,mm.
\subsection{Reduction of MoO$_{3}$ by hydrogen}
In addition to the compression tube and the QMS technique, a supplementary attempt was made to
determine the beam profile by exposing molybdenium trioxide (a yellowish powder)
on a glass plate to the beam. The principle of this method is based on the reduction of MoO$_3$
to a lower oxide of blue colour. It first was used in the experiment to measure the magnetic
moment of the hydrogen atom by splitting of the beam in an inhomogeneous magnetic
field ~\cite{Phipps_Taylor_1927}.

\begin{figure}[hbt]
\includegraphics[width=5cm]{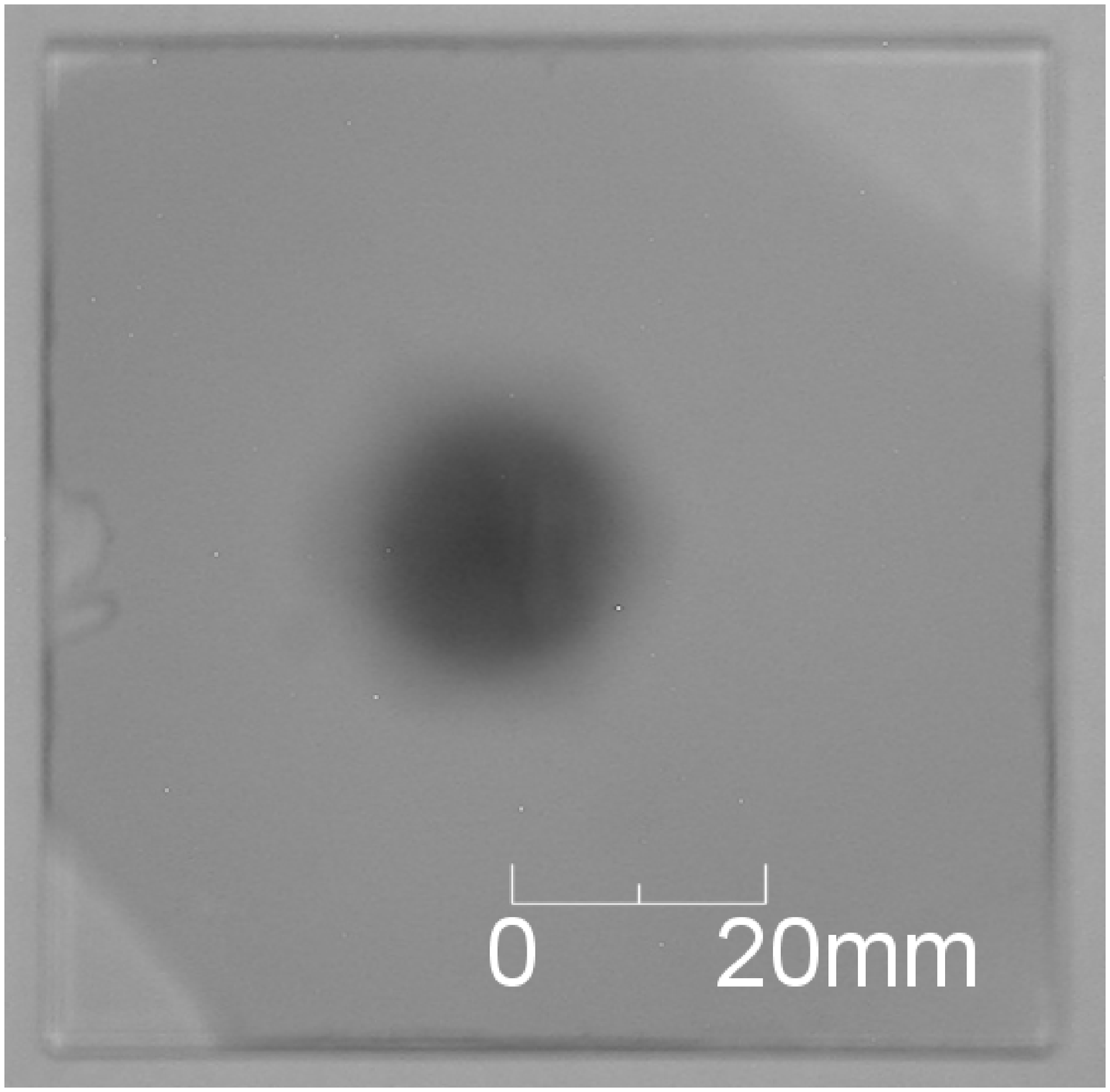}
\caption{Photo of the glass plate covered with molybdenum trioxide $\rm{MoO}_3$ exposed to the
atomic hydrogen beam.}
\label{MoO3}
\end{figure}

This method is much simpler than the time-consuming measurements described in Secs.~\ref{Sec:V-CTprofile} and~\ref{Sec:V-QMSProfile}. It gives qualitative results as presented
in Fig.~\ref{MoO3}. A quantitative analysis, however, requires development of the measuring
technique (e.g., preparation of glass plates, study of the optimum exposure time, digital image
processing).

\subsection{Summary of the profile measurements}
Table~\ref{ProfileTab} summarizes results of the measurements of the ABS beam profile
with the compression-tube and the QMS setup.
The larger errors of the widths, measured with the QMS, are due to the lack of measurements
with the dissociator switched off and the necessity to estimate the background signal from
the existing data. Within the errors, the measured widths do not show a dependence on the distance
from the last magnet. This facilitates to position the feeding tube of the storage cell in a wide
range of a distances to the last magnet. The average values $\Gamma_x=(6.38\pm0.60)$\,mm and
$\Gamma_y=(6.84\pm0.33)$\,mm agree within the errors and yield a common width of
$\Gamma_{x,y}=(6.73\pm0.29)$\,mm. The two-dimensional Gaussian distribution of this width allows
one to estimate the fraction $\eta$ of the beam intensity injected into the compression tube or
a feeding tube. For a tube of 10\,mm diameter $\eta=0.78\pm0.03$, comparable with
$\eta\approx0.7$ given in Sec.~\ref{Sec:V-CTprofile}.
\begin{table}[hbt]
\renewcommand*{\arraystretch}{2}
\caption{Dimensions (FWHM) of the atomic hydrogen beam measured with the compression tube (CT) and
the crossed-beam quadrupole mass spectrometer (QMS) at distances $z$ to the last magnet along
perpendicular directions $x$ and $y$.}
\begin{ruledtabular}
\begin{tabular}{cccc}
      & z[mm] & $\Gamma_{x}$[mm] & $\Gamma_{y}$[mm]\\\hline
CT    & 300   &$6.42\pm 0.09$    &$6.99\pm 0.06$\\
CT    & 337   &$6.27\pm 0.08$    &$6.58\pm 0.08$\\
QMS   & 567   &$7.36\pm 0.43$    &$6.68\pm 0.80$\\
QMS   & 697   &$6.69\pm 0.22$    &$6.38\pm 0.27$\\
\end{tabular}
\end{ruledtabular}
\label{ProfileTab}
\end{table}

\section{Degree of dissociation\label{Sec:VI}}
Besides the intensity of the atomic beam it is important to determine the molecular fraction in
the beam. Molecules injected into the feeding tube, reduce the polarization of the target gas.
\subsection{Measurements with crossed-beam QMS\label{Subsec:AlphaQMS}}
In addition to the data on the profile of the atomic hydrogen beam (Sec.~\ref{Sec:V-QMSProfile}),
data on the distributions of molecular hydrogen in the beam were taken, too, at the positions
$z=567$\,mm and 697\,mm behind the last magnet. The relation between the degree of dissociation
and the QMS signals by the atomic and molecular beam component was given above by
Eq.~(\ref{alpha free jet}). The coefficient $k_{v}=\overline{v}_{m}/\overline{v}_{a}$, however,
is chosen here under the assumption that the average velocity of the atoms is determined by the
nozzle temperature of 65\,K and that of the molecules by scattering and recombination on the ABS
chamber walls at 290\,K. This yields $k_{v}=\sqrt{2\cdot65/290}=0.67$, in good agreement
with Ref.~\cite{Nass_et_al_2003}, where this coefficient was determined by the measured velocity
distributions under similar conditions.

\begin{figure}[b]
\includegraphics[width=\columnwidth]{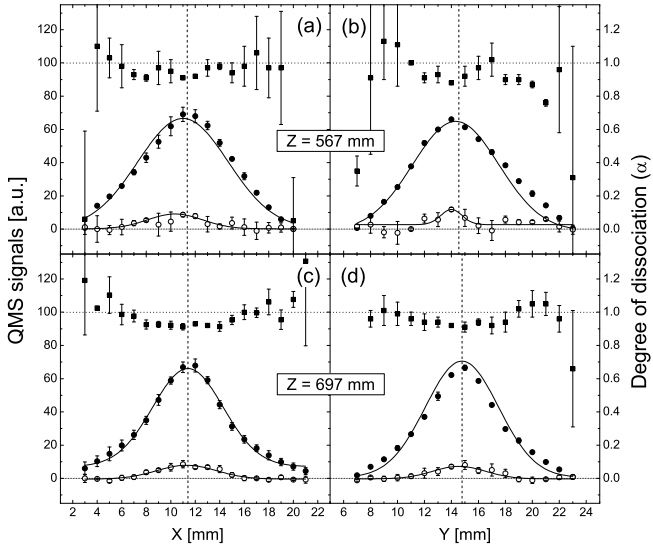}
\caption{Spatial distributions of $\rm{H}_{1}\,(\bullet), \rm{H}_{2}\,(\circ)$ and degree of
dissociation ($^{_\blacksquare}$) averaged over 3\,mm wide bands in the $xz$ and $yz$ planes,
respectively (here the $z$-axis is the geometrical axis of the ABS).}
\label{DOD-profile1}
\end{figure}

The measured profiles of the atomic fraction (identical to those of Fig.~\ref{Profile2}), those of
the molecular fraction, and those of the degree of dissociation, deduced from
Eq.~(\ref{alpha free jet}), are collected in Fig.~\ref{DOD-profile1}.

As it is seen from the figure, the distribution of the degree of dissociation shows a
dip around the central line due to the higher density of molecular hydrogen originating from
the nozzle. The mean value in an aperture of 10\,mm diameter results as
$\overline{\alpha}\,=\,0.95\,\pm\,0.04$.
\subsection{Measurements with the Lamb-shift polarimeter}
A cup in the quench chamber of the Lamb-shift polarimeter (LSP), described in
Ref.~\cite{Engels_et_al_2005_1}, allows one to measure the currents $I_{\rm cup}{\rm(H_{1})}$ and
$I_{\rm cup}{\rm(H_{2})}$ of the $\rm{H}_{1}^{+}$ and $\rm{H}_{2}^{+}$ ions, extracted from the
ionizer and separated by the Wien filter with the cesium evaporation and the spin filter switched
off. The relation between the degree of dissociation $\alpha$ and the measured currents is
\begin{equation}
\alpha=\frac{I_{\rm cup}({\rm H_1})-\frac{r_{1}}{r_{2}}I_{\rm cup}({\rm H_2})}
            {I_{\rm cup}({\rm H_1})-\frac{r_{1}}{r_{2}}I_{\rm cup}({\rm H_2})
         +2\frac{k_{v}}{r_{2}}I_{\rm cup}({\rm H_2})}.
\label{alpha-LSP}
\end{equation}
Among the three coefficients, $k_{v}=0.67$ as for the measurement with the QMS. For the electron
energy of about 100\,keV, the ratio $r_{1}$ of dissociative to non-dissociative ionization of
${\rm H_{2}}$ is~\cite{Engels_et_al_2005_1}
\begin{equation}
r_{1}=\frac{\sigma({\rm H_{2}}\rightarrow2{\rm H_{1}^{+}})}{\sigma({\rm H_{2}}\rightarrow {\rm
H_{2}^{+}})}=0.095\pm 0.008
\end{equation}
and the ratio between the ionization cross sections is~\cite{ionization CS,Engels_et_al_2005_1}
\begin{equation}
r_{2}=\frac{\sigma_{\rm ion}({\rm H_{2}})}{\sigma_{\rm ion}({\rm H_{1}})}=1.7\pm 0.1.
\end{equation}

At the standard operation parameters of the source (Sec.~\ref{Sec:IV}), the measured currents are
$I_{\rm cup}({\rm H_{1}})=(125\pm 0.5)$\,nA and $I_{\rm cup}({\rm H_{2}})=(6.4\pm 0.1)$\,nA,
yielding $\bar{\alpha}=(0.96\pm 0.04)$, in excellent agreement with the value resulting from
the measurements with the QMS (Sec.~\ref{Subsec:AlphaQMS}).

\section{Beam polarization\label{Sec:VII}}
The Lamb-shift polarimeter was designed, built, and tested at Universit\"at zu
K\"oln~\cite{Engels_et_al_2003}. It was used to measure and to optimize the polarization of
the atomic hydrogen and deuterium beams delivered by the ABS. Details are found in
Ref.~\cite{Engels_et_al_2003}.

The vector polarization $p_{z}$ for hydrogen is defined by the relative hyperfine-state
occupation numbers $N(m_I)$,
\begin{equation}
p_z=\frac{N(+\frac{1}{2})-N(-\frac{1}{2})}{N(+\frac{1}{2})+N(-\frac{1}{2})}\,,
\label{+PzHexp}
\end{equation}
for deuterium
\begin{equation}
p_z=\frac{N(+1)-N(-1)}{N(+1)+N(0)+N(-1)}\,.
\end{equation}
Deuterium tensor polarization $p_{zz}$ is given by
\begin{equation}
p_{zz}=\frac{N(+1)+N(-1)-2\cdot N(0)}{N(+1)+N(0)+N(-1)}\,.
\end{equation}
These polarizations can be derived from the measured Lyman-$\alpha$ peak strengths $S$ by
application of a number of correction factors~\cite{Engels_et_al_2003, Engels_et_al_2005_1}.

\begin{figure}[hbt]
\includegraphics[width=8.5cm, angle=0]{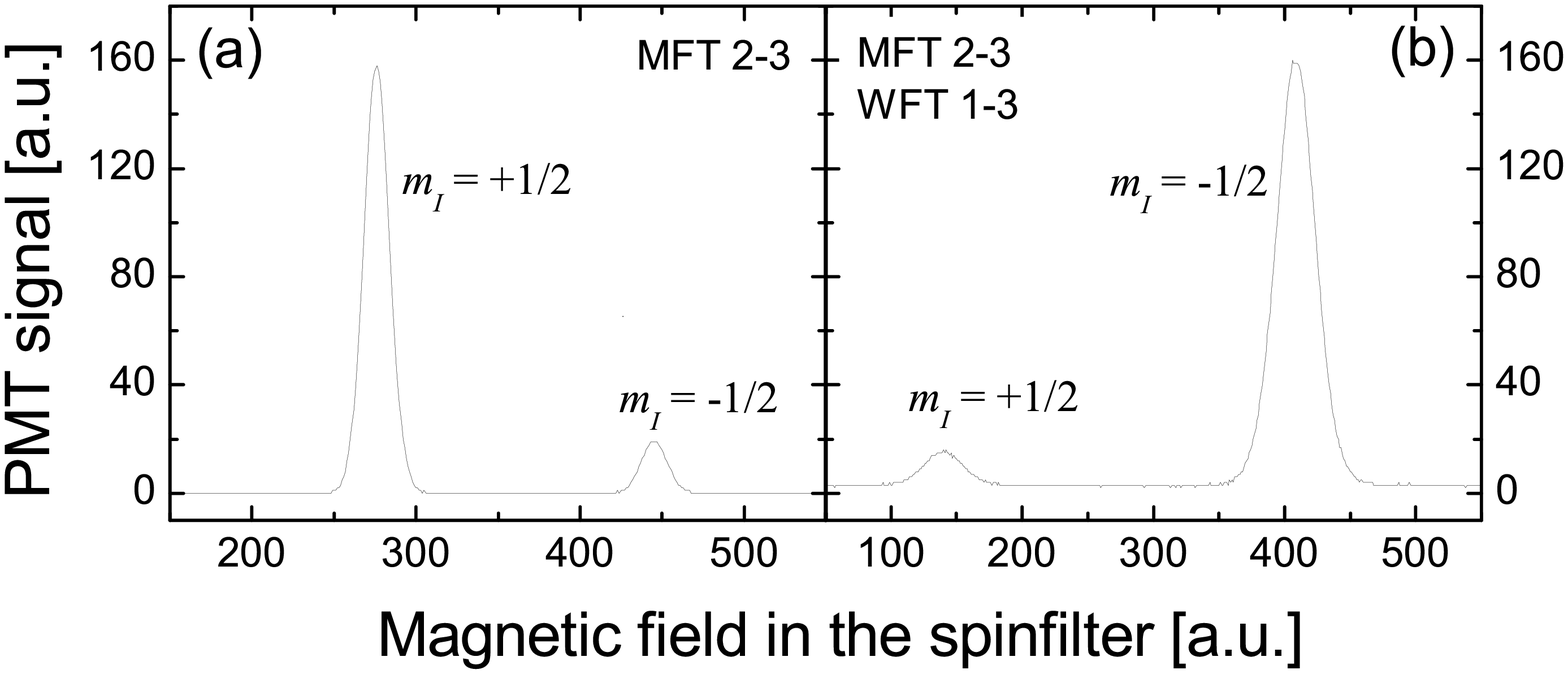}
\caption{Lyman-$\alpha$ spectra measured with the polarized hydrogen beam. (a): population change
from state $|2\rangle$ to state $|3\rangle$ induced by the MFT unit; (b) same as (a) with
subsequent population change from state $|1\rangle$ to state $|3\rangle$
induced by the WFT unit.}
\label{Lyman-hydrogen}
\end{figure}

\begin{figure}[hbt]
\includegraphics[width=8.5cm, angle=0]{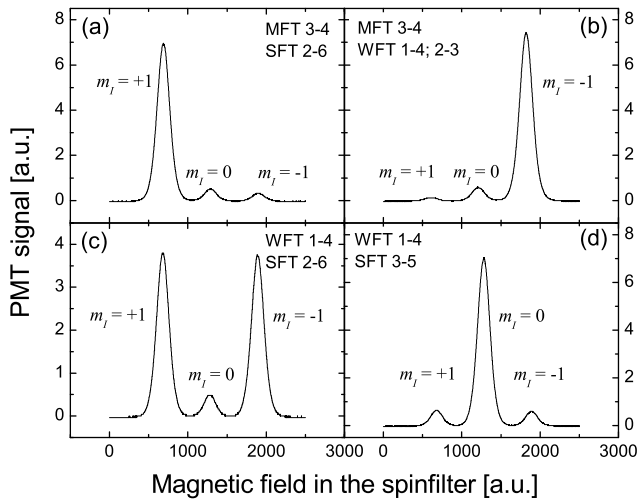}
\caption{Lyman-$\alpha$ spectra measured with the polarized deuterium beam. (a) and (b) vector
polarization resulting from subsequent transitions MFT ($3\rightarrow4$) and SFT ($2\rightarrow6$)
and WFT ($1\rightarrow4, 2\rightarrow3$), respectively; (c) and (d) tensor polarization resulting
from subsequent transitions WFT ($1\rightarrow4$) and SFT ($2\rightarrow6$) and SFT
($3\rightarrow5$), respectively.}
\label{Lyman-deuterium}
\end{figure}

Typical Lyman $\alpha$ spectra, measured with the polarized hydrogen and deuterium beam from the
ABS, are shown in the Figs.~\ref{Lyman-hydrogen} and~\ref{Lyman-deuterium}.

\begin{table}[h]
\caption{The vector polarization $p_{z}$ of the hydrogen beam and the vector and the
tensor polarization $p_{zz}$ of the deuterium beam from the ABS measured with the Lamb-shift
polarimeter.}
\renewcommand*{\arraystretch}{1.4}
\begin{ruledtabular}
\begin{tabular}{cccc}
          & populated                     & $p_{z}$           & $p_{zz}$\\
          & state(s)                      &                   &\\
\hline
Hydrogen  & $|1\rangle$               & $+0.89\pm 0.01$   & -\\\vspace{2mm}
          & $|3\rangle$               & $-0.96\pm 0.01$   & -\\
Deuterium & $|1\rangle+|6\rangle$ & $+0.88\pm 0.01$   & $+0.88\pm 0.03$\\
          & $|3\rangle+|4\rangle$ & $-0.91\pm 0.01$   & $+0.85\pm 0.02$\\
          & $|3\rangle+|6\rangle$ & $+0.005\pm 0.003$ & $+0.90\pm 0.01$\\
          & $|2\rangle+|5\rangle$ & $+0.005\pm 0.003$ & $-1.71\pm 0.03$\\
\end{tabular}
\end{ruledtabular}
\label{Polexp}
\end{table}

The polarization values for the hydrogen and the deuterium beam, derived from the Lyman-$\alpha$
peak-strength ratios with  application of the necessary corrections, are summarized in
Table~\ref{Polexp}.

The vector polarization for hydrogen of the first line reflects the population of state $|1\rangle$
and state $|2\rangle$ according to the Eqs.~(\ref{+PzHest}) and~(\ref{+PzHexp}). The value of 0.91,
deduced from the calculated transmission values, is confirmed by the measured one.

\section{Conclusions and Outlook\label{Sec:VIII}}
In this paper, we present the detailed description of the major components of the atomic beam
source (ABS) for the polarized internal gas target of the magnet spectrometer ANKE in
COSY-J\"ulich. The ABS was built for the purpose of extending the physics program of ANKE from
unpolarized and single-polarized investigations with stored beams towards double-polarized
experiments~\cite{COSY152}, thus facilitating nuclear reaction studies involving $\vec p\vec p$, $\vec p\vec d$, $\vec d\vec p$ and $\vec d\vec d$ initial states.

The mechanical design took into account that at ANKE the source has to be mounted vertically and
transversely movable together with the transverse motion of the spectrometer magnet D2. The
design of the system of sextupole magnets took advantage of the developments in the field
of rare-earth permanent magnets (NdFeB). Dedicated tools and methods were developed to determine
and to optimize the source parameters, i.e., intensity, degree of dissociation, and polarization.
Special emphasis was put on the measurements of the spatial distributions of the atomic and
molecular beam near the focus, where the feeding tube of the storage cell is located. The ABS has
been used in a number of investigations at ANKE, the commissioning effort to prepare the target
for the use with polarized H is described in Ref.~\cite{Max_et_al_2011}. Performed studies of
the deuteron-charge exchange reaction are summarized in
Ref.~\cite{David M_et_al_2011,Rathmann_2011}, studies in near-threshold pion production are
reported about in Ref.~\cite{Sergej D_2011}.

The ABS resides at the ANKE target position for a few months per year only, thus during the
remaining time it is used for other studies. It had been observed that the nuclear
polarization in recombined hydrogen is partially retained after
recombination~\cite{Wise_et_al_2001}, as well as evidence for nuclear tensor polarization in
recombined deuterium molecules~\cite{v_d_Brand_et_al_1997}. In order to investigate this
recombination process in more detail, a special setup has been developed in the framework of an
ISTC project~\cite{ISTC_2001}, and the recombination process for different cell-wall coatings, and
different polarizations of the injected hydrogen or deuterium atoms as function of cell-wall
temperature, strength of the magnetic holding field, and gas pressure in the cell is presently
investigated~\cite{Engels_et_al_2010,DFG,Engels_et_al_2011}.

\vspace*{0.5cm}
\appendix
\section{Preparation of Discharge Tubes and Nozzles\label{Sec:Appendix}}
\label{sect:appA}
\subsection{Tube Treatment}
One end of the discharge tube is machined at a 45$^\circ$ angle, while the other is kept flat.
Both ends are then remelted and the tubes are tempered at 150\,$^\circ$C. The tubes are further
treated according to the procedure described in Ref.~\cite{Koch_Diss_1999}, which includes
successive cleaning with acetone, methanol, distilled water, and subsequent rinsing by a 2:1
acid mixture of concentrated HF (40\%) and HCl~(32\%) for 5~min.  The tubes are then flushed
by distilled water and dried.
\subsection{Nozzle Treatment}
The nozzles are cleaned in an ultrasonic bath of trichlorethylene, acetone, methanol, and finally
distilled water, all at 50\,$^\circ$C. Anodizing takes place in sulfuric acid to form a thin layer
of $\rm Al_2O_3$, as described in Ref.~\cite{Koch_Diss_1999}. Afterwards they are immersed in
distilled water for 30\,min at 95\,$^\circ$C.

\begin{acknowledgments}
The authors want to thank O.W.B.\,Schult, Institut f\"ur Kernphysik (IKP),
J\"ulich, who initiated the polarization program of ANKE. Thanks go to the
design office, the mechanical workshop, and especially to W.R.\,Ermer, all IKP. Valuable advice
was received from the PINTEX collaboration at IUCF, from the target
group at HERMES, especially N.\,Koch, and from D.\,Toporkov,
BINP, Novosibirsk. The support by V.\,Koptev, PNPI, Gatchina, who regrettably passed away in
January 2012, is gratefully acknowledged. Thanks go also to R.\,Poprawe and colleagues,
Fraunhofer-Institut f\"ur Lasertechnik, Aachen, where the encapsulations of the magnets were
laser-welded.
\end{acknowledgments}
\pagebreak



\begin{thebibliography}{9}
\bibitem{COSY152} A.\,Kacharava, F.\,Rathmann, and C.\,Wilkin, Spin Physics from COSY to FAIR,
COSY Experiment Proposal No. {\bf 152} (2005).\\
Available under \url{http://arXiv:nucl-ex/0511028}.
\bibitem{Haeberli_1965} W.\,Haeberli, in Proc. $2^{nd}$ Int. Symp. on Polarization Phenomena of
Nucleons, Karlsruhe 1965. Eds. P. Huber and H. Schopper, Experientia Supplementum {\bf 12}, 64
(Birkh\"auser Verlag, 1966).
\bibitem{Steffens+Haeberli_2003} E.\,Steffens and W.\,Haeberli, Rep. Progr. Phys. {\bf 66}, 1887
(2003).
\bibitem{Barsov_et_al_2001} S.\,Barsov et al., Nucl. Instr. and Meth. A {\bf 462}, 364 (2001).
\bibitem{Maier_1997} R.\,Maier, Nucl. Instr. and Meth. A {\bf 390}, 1 (1997).
\bibitem{KG_2012} K.\,Grigoryev et al., Proc. 14th International Workshop on Polarized Sources,
Targets and Polarimetry (PSTP 2011), 12-16 September 2011, St.Petersburg, Russia, eds.
K.\,Grigoryev, P.\,Kravtsov and A.\,Vasilyev, ISBN 978-5-86763-282-3, 61 (2011).
\bibitem{Engels_et_al_2003} R.\,Engels et al., Rev. Sci. Instrum. {\bf 74}, 4607 (2003).
\bibitem{Engels_et_al_2005_1} R.\,Engels et al., Rev. Sci. Instrum. {\bf 76}, 053305 (2005).
\bibitem{Wise_et_al_1993} T.\,Wise et al., Nucl. Instr. and Meth. A {\bf 336}, 410 (1993).
\bibitem{Dezarn_et_al_1995} W.A.\,Dezarn et al., Nucl. Instr. and Meth. A {\bf 362}, 36 (1995).
\bibitem{Rinckel et al 2000}T.\, Rinckel et al., Nucl. Instr. and Meth. A {\bf 439}, 117 (2000).
\bibitem{Stock_et_al_1994} F.\,Stock et al., Nucl. Instr. and Meth. A {\bf 343}, 334 (1994).
\bibitem{Nass_et_al_2003} A.\,Nass et al., Nucl. Instr. and Meth. A {\bf 505}, 633 (2003).
\bibitem{Derenchuk_et_al_1994} V.\,Derenchuk et al., Proc. Conf. Polarized Ion Sources and
Polarized Gas Targets, Madison, WI, 1993. Eds. L.W.\,Anderson and W.\,Haeberli, AIP Conf.
Proc. {\bf 293}, 72 (American Institute of Physics, 1994).
\bibitem{Okamura_et_al_1994} H.\,Okamura et al., see Ref.~\cite{Derenchuk_et_al_1994}, p. 84.
\bibitem{Hatanaka_et_al_1997} K.\,Hatanaka et al., Nucl. Instr. and Meth. A {\bf 384}, 575 (1997).
\bibitem{Schiffer} Manufacturer Schiffer Metall- \& Vakuumtechnik, 52428 J\"ulich, Germany.
\bibitem{RGS120} Single-stage type RGS/120, refrigerating capacity 120\,W at 80\,K and 20\,W
at 30\,K, Leybold Vacuum GmbH, 50968 K\"oln, Germany.
\bibitem{VAT-mini} Mini UHV gate valve, series 010, VAT Germany GmbH, 85630 Grasbrunn,
Germany.
\bibitem{Fomblin} Type F3 fomblin oil, Pfeiffer Vacuum GmbH, 35614 Asslar, Germany.
\bibitem{HU1} Model HU 1, Leybold Vacuum GmbH, 50968 K\"oln, Germany.
\bibitem{SKI} Manufacturer SK Industriemodell GmbH, 52531 \"Ubach-Palenberg, Germany.
\bibitem{PFG600} Type PFG 600 RF with automatic matchbox PFM 1500 A-IND, H\"uttinger Elektronik
GmbH, 79110 Freiburg, Germany.
\bibitem{Korsch_Diss_1990} W.\,Korsch, PhD Thesis, Philipps Universit\"at Marburg (1990).
\bibitem{Stock_et_al_1996} F.\,Stock et al., Int. Workshop on Polarized Beams and Polarized Gas
Targets, Koeln, Germany, 1995. Eds. H.\,Paetz\,gen.\,Schieck and L.\,Sydow (World Scientific
Publ. Co., 1996) p. 260.
\bibitem{TubeDia} The first number denotes the outer diameter and the second one the wall
thickness.
\bibitem{DURAN} Type Duran 8330, equivalent to Corning 7740 (Pyrex), Schott AG,
55122 Mainz, Germany.
\bibitem{LAUDA} Ultra-Kryomat RUL 80-D, Lauda Dr.\,R.\,Wobser GmbH, 97912 Lauda-K\"onigshofen,
Germany.
\bibitem{ODU} ODU-Kontakt GmbH, 84444 M\"uhldorf, Germany.
\bibitem{Handbuch_Chem_Phys} Handbook of Chemistry and Physics, Ed. R.C.\,East (The Chemical
Rubber Co., 1973), p. E-10.
\bibitem{Koch+Steffens_1999} N.\,Koch and E.\,Steffens, Rev.\,Sci.\,Instrum. {\bf 70}, 1631
(1999).
\bibitem{Vassiliev_et_al_1997} A.\,Vassiliev et al., Petersburg Nuclear Physics Institute Report
NP-32-1997 No. 2175 (1997).
\bibitem{Lorentz_Dipl_1993} B.\,Lorentz, Diploma Thesis, Ruprecht-Karls-Universit\"at Heidelberg
(1993).
\bibitem{Nass+Steffens_2009} A.\,Nass and E.\,Steffens, Nucl. Instr. and Meth. A {\bf 598}, 653
(2009).
\bibitem{Haeberli_1967} W.\,Haeberli, Ann. Rev. Nucl. Sci. {\bf 17}, 373 (1967).
\bibitem{Vassiliev_et_al_2000} A.\,Vassiliev et al., Rev. Sci. Instr., {\bf 71}, 3331 (2000).
\bibitem{Kubischta_1991} W.\,Kubischta, Proc. Workshop on Polarized Gas Targets for Storage Rings,
Heidelberg, 23-26 September 1991, Eds. H.G.\,Gaul, E.\,Steffens, and K.\,Zapfe
(Max-Planck-Institut f\"ur Kernphysik Heidelberg).
\bibitem{HFS} The labeling of the hyperfine states as
$|1\rangle=|m_{j}=+1/2,m_{I}=+1/2\rangle$, $|2\rangle=|+1/2,-1/2\rangle$,
$|3\rangle=|-1/2,+1/2\rangle$, and $|4\rangle=|-1/2,-1/2\rangle$ for hydrogen and $|1\rangle
=|+1/2,+1\rangle$, $|2\rangle =|+1/2,0\rangle$, $|3\rangle
=|+1/2,-1\rangle$, $|4\rangle =|-1/2,-1\rangle$, $|5\rangle =|-1/2,0\rangle$, and $|6\rangle
=|-1/2,+1\rangle$ for deuterium follows that of Ref.~\cite{Haeberli_1967}.
\bibitem{VACODYM} Produced from VACODYM 510HR, 383HR, and 400HR by Vacu\-umschmelze GmbH, 63412
Hanau, Germany.
\bibitem{Halbach_1980} K.\,Halbach, Nucl. Instr. and Meth. {\bf 169}, 1 (1980).
\bibitem{FIL} Welding performed at Fraunhofer-Institut f\"ur Lasertechnik, 52074 Aachen, Germany.
\bibitem{Abragam+Winter_1958} A.\,Abragam and J.M.~Winter, Phys. Rev. Lett. {\bf 1}, 374 (1958).
\bibitem{Lorenz_Dipl_1999} S.\,Lorenz, Diploma Thesis, Friedrich-Alexander-Universit\"at
Erlangen-N\"urnberg (1999).
\bibitem{Gaul+Steffens_1992} H.-G.\,Gaul and E.\,Steffens, Nucl. Instr. and Meth. A {\bf 316}, 297
(1992).
\bibitem{Oh_1970} S.\,Oh, Nucl. Instr. and Meth. {\bf 82}, 189 (1970).
\bibitem{Schieck_2008} H.\,Paetz gen. Schieck, Nucl. Instr. and Meth. A {\bf 587}, 213 (2008).
\bibitem{Philpott_1987} R.J.\,Philpott, Nucl. Instr. and Meth. A {\bf 259}, 317 (1987).
\bibitem{Jaensch_et_al_1985} H.\,J\"ansch et al., Hyperfine Interactions {\bf 22}, 253 (1985).
\bibitem{Roberts_et_al_1992} A.D.\,Roberts et al., Nucl. Instr. and Meth. A {\bf 322}, 6 (1992).
\bibitem{Capiluppi_2012} M.\,Capiluppi et al.,
\url{http://theor.jinr.ru/~spin2012/talks/s6/Steffens.pdf} (to be published in Physics of
Elementary Particles and Atomic Nuclei, JINR, Russia,
\url{http://pepan.jinr.ru/pepan/eng/about/}).
\bibitem{PNPI} Manufactured by St.\,Petersburg Nuclear Physics Institute, 188300 Gatchina, Russia.
\bibitem{Kleines_et_al_2006}H.\,Kleines et al., Nucl. Instr. Meth. A {\bf 560}, 503 (2006).
\bibitem{Vassiliev_et_al_1998} A.\,Vassiliev et al., Petersburg Nuclear Physics Institute Report
EP-46-1998 No. 2260 (1998).
\bibitem{Vassiliev_et_al_1999} A.\,Vassiliev et al., Proc. Int. Workshop Polarized Sources and
Targets, Erlangen, Germany,
September 29 -October 2, 1999. Eds. A.\,Gute, S.\,Lorenz, E.\,Steffens (Universit\"at
Erlangen-N\"urnberg, 1999), p.~200.
\bibitem{Max_Dipl_1999} M.\,Mikirtytchiants, Diploma Thesis, St. Petersburg State Technical
University (1999).
\bibitem{Max_et_al_1999} M.\,Mikirtytchiants et al., see Ref.~\cite{Vassiliev_et_al_1999}, p.~478.
\bibitem{ionization CS} Y.K.\,Kim et al., Electron-impact cross section database, 2002,\\
\url{http://pysics.nist.gov/PhysRefData/Ionization}.
\bibitem{Nekipelov1} M.\,Nekipelov, Diploma Thesis, St. Petersburg State Technical University
(1999).
\bibitem{Nekipelov2} M.\,Nekipelov et al., see Ref.~\cite{Vassiliev_et_al_1999}, p.~486.
\bibitem{Roth_Vakuum} A.\,Roth, Vacuum Technology (Elsevier, Amsterdam, 1996).
\bibitem{Phipps_Taylor_1927} T.E.\,Phipps and J.B.\,Taylor, Phys. Rev. {\bf 29}, 309 (1927).
\bibitem{Max_et_al_2011} M.\,Mikirtychyants et al., J. Phys.: Conf. Ser. 295, 012148 (2011).
\bibitem{David M_et_al_2011} D.\,Mchedlishvili et al., J. Phys.: Conf. Ser. 295, 012099 (2011).
\bibitem{Rathmann_2011} F.\,Rathmann, J. Phys.: Conf. Ser. 295, 012006 (2011).
\bibitem{Sergej D_2011} S.\,Dymov (for the ANKE collaboration), J. Phys.: Conf. Ser. 295, 012095
(2011).
\bibitem{Wise_et_al_2001} T.\,Wise et al., Phys.\,Rev.\,Lett. {\bf 87}, 042701 (2001).
\bibitem{v_d_Brand_et_al_1997} J.F.J.\,van\,den\,Brand et al., Phys.\,Rev.\,Lett. {\bf 78}, 1235
(1997).
\bibitem{ISTC_2001} International Science and Technology Center, Project No. 1861.
\bibitem{DFG} Work now financed by Deutsche Forschungsgemeinschaft, project 436 RUS 113/977/01.
\bibitem{Engels_et_al_2010} R.\,Engels et al., Proc. 13$^{\rm th}$ Int. Workshop on Polarized
Sources, Targets and Polarimetry, Ferrara, Italy, September 7-11, 2009. Eds. G.\,Ciullo,
M.\,Contalbrigo, P.\,Lenisa (World Scientific, 2011), p.~215.
\bibitem{Engels_et_al_2011} R.\,Engels et al., J. Phys.: Conf. Ser. 295, 012161 (2011).
\bibitem{Koch_Diss_1999} N.\,Koch, PhD Thesis, Friedrich-Alexander-Universit\"at
Erlangen-N\"urnberg (1999).

\end{thebibliography}
\end{document}